\documentclass[11pt,reqno]{amsart}

\usepackage{amsmath,amsthm,amssymb,amsfonts,mathtools}
\usepackage{bm}
\usepackage{natbib}
\usepackage{bbm}
\usepackage{graphicx}
\usepackage{multirow}
\usepackage{here}
\usepackage{fullpage}
\usepackage{xurl}
\usepackage{color}
\usepackage{xcolor}
\usepackage{mathrsfs}
\usepackage{enumitem}
\usepackage{comment}
\usepackage{subcaption}
\usepackage{pifont}
\usepackage[linesnumbered,ruled,vlined]{algorithm2e}
\usepackage[hidelinks]{hyperref}

\makeatletter
\makeatletter

\newtheorem{theorem}{Theorem}[section]
\newtheorem{corollary}{Corollary}[section]
\newtheorem{lemma}{Lemma}[section]
\newtheorem{proposition}{Proposition}[section]
\newtheorem{assumption}{Assumption}[section]
\theoremstyle{definition}
\newtheorem{definition}{Definition}[section]
\newtheorem{remark}{Remark}[section]
\newtheorem{example}{Example}[section]

\def\E#1{\mathbb{E}\left[#1\right]}
\def\P#1{\mathbb{P}\left[#1\right]}

\newcommand{\argmin}{\mathop{\rm arg~min}\limits}

\renewcommand{\tilde}{\widetilde}
\renewcommand{\hat}{\widehat}

\def\f{Fr\'echet }
\def\bco{\iffalse} 
\allowdisplaybreaks

\makeatother

\begin{document}

\title[]{Lee bounds for random objects}
\thanks{} 

\author[D. Kurisu]{Daisuke Kurisu}
\author[Y. Okamoto]{Yuta Okamoto}
\author[T. Otsu]{Taisuke Otsu}

\date{First version: \today}

\address[D. Kurisu]{Center for Spatial Information Science, The University of Tokyo, 5-1-5, Kashiwanoha, Kashiwa-shi, Chiba 277-8568, Japan.}
\email{daisukekurisu@csis.u-tokyo.ac.jp}

\address[Y. Okamoto]{Graduate School of Economics, Kyoto University, Yoshida Honmachi, Sakyo, Kyoto 606-8501, Japan.
}
\email{okamoto.yuuta.57w@st.kyoto-u.ac.jp}

\address[T. Otsu]{Department of Economics, London School of Economics, Houghton Street, London, WC2A 2AE, UK.
}
\email{t.otsu@lse.ac.uk}

\begin{abstract}
In applied research, Lee (2009) bounds are widely applied to bound the average treatment effect in the presence of selection bias. This paper extends the methodology of Lee bounds to accommodate outcomes in a general metric space, such as compositional and distributional data. By exploiting a representation of the \f mean of the potential outcome via embedding in an Euclidean or Hilbert space, we present a feasible characterization of the identified set of the causal effect of interest, and then propose its analog estimator and bootstrap confidence region. The proposed method is illustrated by numerical examples on compositional and distributional data.
\end{abstract}

\keywords{}
\maketitle

\section{Introduction}\label{sec:intro}

Randomized controlled trials are among the most credible approaches for answering causal questions in economics, medical science, and many other disciplines. However, their validity can be undermined by endogenous sample attrition or non-response, which are ubiquitous in empirical studies. A common approach to address this issue, developed in a seminal work by \cite{Lee:2009}, is to bound the average treatment effect of units whose outcomes are observed regardless of the treatment status. Recent empirical studies employing the Lee bounds include \cite{Alfonsi_etal:2020}, \cite{Atal_etal:2024}, and \cite{Dobbie_Fryer:2015}, among many others.

The methodology of the Lee bounds has been extended in several directions to deal with noncompliance issues \citep{Chen_Flores:2015}, continuous treatments \citep{Lee_liu:2025}, multi-layered selection \citep{Kroft_etal:2025}, and relaxed monotonicity assumptions as well as multiple outcomes \citep{Semenova:2025}, for example. However, almost all existing works focus on outcomes that lie in a Euclidean space, even though many economically important outcomes do not live in Euclidean spaces.

One prominent class is compositional data, where components are nonnegative and sum to one. For example, labor and macro economists study intra-household time allocation (e.g., \citealp{Cardia_Gomme:2018, Lise_Yamada:2019}), while public economists analyze the composition of government expenditures (e.g., \citealp{Brender_Drazen:2013, Shelton:2007}). Other relevant examples include the composition of household consumption, energy mix across sources, agricultural land use, and portfolio shares in finance.

Another important class consists of distribution-valued outcomes. Economists have studied income and wage distributions (e.g., \citealp{Battistin_etal:2009, DiNardo_etal:1996}) as well as price dispersion (e.g., \citealp{Cavallo:2017, Pennerstorfer_etal:2020}). Related examples include the distributions of daily step counts, daily television viewing time, or hourly electricity consumption, which may be of interest in health, education, and public economics.

Beyond these classes, researchers also encounter network outcomes, such as cross-country trade and travel flows, as well as transaction networks among firms. Interval-valued data are likewise ubiquitous in econometric analysis---for example, when outcomes are reported in brackets for confidentiality, elicited as ranges to improve response rates, or observed with rounding error. Finally, correlation matrices are central objects for studying connectivity, including applications such as product recommendation and brain functional connectivity. We refer to \cite{Dubey:2024} and references therein for a general review of statistical analysis of such random objects, along with a range of applications.

Despite their prevalence, standard implementations of Lee bounds are not directly applicable to these cases, since such outcomes are not situated in Euclidean spaces. To overcome this limitation, we extend the scope of partial identification analysis via Lee bounds by adapting the methodology of metric statistics for random objects. In particular, we define the expected potential outcomes for random objects by introducing the notion of the \f mean, a direct generalization of the conventional mean toward a general metric space,\footnote{Perhaps a popular example of the \f mean in economics is the Aumann mean for random sets (e.g., \citealp{BeMo08}), which is shown to be a special case of the \f mean (\citealp{kuri:25}), and our methodology also applies to random sets (see, Example \ref{ex:int} below).} and then establish partial identification results on the expected potential outcome by exploiting a representation in an embedded Euclidean or Hilbert space, where the bounding argument in \cite{Lee:2009} may be adapted. Intuitively, our identified set is obtained by pulling back the Lee-type bounds on the embedded Euclidean or Hilbertian variables, and this approach enables analysis that takes into account the geometric structure of the metric space in which the outcome takes values. Based on our partial identification strategy, the treatment effect can be characterized by a difference in the embedded space, a difference in the projections, or geodesics in the metric space. Our estimator for the identified set is obtained by taking the sample analogs, and we present a valid bootstrap procedure for the confidence region of our identified set.

Causal inference on random object outcomes has been increasingly popular in recent econometrics and statistics literature, e.g., \cite{Gunsilius:2023}, \cite{kurisu_etal:24, KuZhOtMu25a, KuZhOtMu25b}, and \cite{ZhKuOtMu25}. The present paper contributes to this literature by extending the method of Lee bounds to random objects. In contrast to the above papers, to the best of our knowledge, this is the first paper that conducts partial identification analysis on the \f mean in a general metric space. In this sense, this paper opens a new avenue for applications of the methodologies of partial identification and metric statistics.

This paper is organized as follows. After closing this section with a recap of the conventional Lee bounds, Section \ref{sec:bench} showcases our partial identification analysis by focusing on compositional data. Then Section \ref{sec:gen} presents our general methodology for partial identification. In Section \ref{sec:implementation}, we discuss estimation and bootstrap inference on the proposed identified set. Section \ref{sec:empir} provides an empirical illustration. Finally, Appendix \ref{app:ex} discusses additional examples of random objects, and Appendix \ref{app:pf} contains the proofs of the theoretical results.

\subsection*{Recap: Lee bounds}
For the reader's convenience, we start with the standard Lee bounds for a scalar outcome. We also discuss potential issues that arise and are resolved in the subsequent sections. Consider a binary treatment $D\in \{0,1\}$ (one for the treatment and zero otherwise) with potential outcomes $Y_0$ and $Y_1$ that take values in $\mathcal{Y}\subset\mathbb{R}$. Let $S_{0}, S_{1}\in\{0,1\}$ denote potential selection indicators (one for non-missing and zero otherwise). As in \cite{Lee:2009}, we assume that $D$ is independent from $(Y_1, Y_0, S_1, S_0)$ (random assignment) and $S_1 \ge S_0$ (monotonicity). Instead of the latent variables $(Y_1, Y_0, S_1, S_0)$, we observe $D$ and $(S, Y)$ that are generated from the general sample selection model:
\begin{equation}\label{eq:obs}
\begin{aligned}
    S &= S_{1} D + S_{0} (1-D),\\
    Y &= \begin{cases}
        Y_{1}D + Y_{0} (1-D) & \text{ if } S=1,\\
        \text{NA} & \text{ if } S=0.
    \end{cases}
\end{aligned}
\end{equation}
When the outcome variable $Y$ is scalar-valued, we typically partially identify the average treatment effect for a certain subpopulation \citep{Lee:2009}:
\begin{align*}
    \E{Y_{1} - Y_{0} \mid S_{1} = S_{0} = 1}
    = 
    \E{Y_{1} \mid S_{1} = S_{0} = 1} - 
    \E{Y_{0} \mid S_{1} = S_{0} = 1},
\end{align*}
where we refer to the group of individuals with $S_{1} = S_{0} = 1$ as the \textit{always-observed} group. The second term $\E{Y_{0} \mid S_{1} = S_{0} = 1}$ is point identified as $\E{Y \mid S = 1, D=0}$, but the first term $\E{Y_{1} \mid S_{1} = S_{0} = 1}$ is not. To partially identify the first term, observe that 
\begin{align*}
    \underbrace{\P{Y\leq y \mid S=1, D=1}}_{\text{identified}}
    =
     p \underbrace{\P{Y_{1}\leq y \mid S_{0}=1, S_{1}=1}}_{\text{always-observed group}} + (1-p)\P{Y_{1}\leq y \mid S_{0}=0, S_{1}=1},
\end{align*}
where $p=\P{S=1 \mid  D=0}/\P{S=1 \mid  D=1}$. Then, by considering the worst (resp. best) case scenario where the smallest (resp. largest) $100\times p\%$ values of observed $Y$ (conditional on $S=1, D=1$) are entirely attributed to $Y_{1}$ of the always-observed group, we can obtain the lower (resp. upper) bound on the first term, leading to the following bounds:
\begin{align*}
    \left[
    \E{Y \mid S=1, D=1, Y \leq Q_Y(p)}, \E{Y \mid S=1, D=1, Y \geq Q_Y(1-p)}
    \right],
\end{align*}
where $Q_{Y}(\cdot)$ denotes the conditional quantile function of $Y$ given $D = 1, S = 1$. Based on these bounds, we can obtain the Lee bounds on the average treatment effect $\E{Y_{1} - Y_{0} \mid S_{1} = S_{0} = 1}$ for the always-observed group.

In the subsequent sections, we extend this partial identification strategy to outcomes that take values in more general metric spaces (i.e., random objects rather than Euclidean variables). One of the key issues is how to define and interpret a ``mean'' in such settings. For Euclidean outcomes, the usual expectation admits a geometric interpretation as the ``center'' of the distribution. This interpretation, however, is inherently metric-dependent, and as a result, when the outcome variable lies in a non-Euclidean space, the conventional Euclidean mean may ignore the geometric property and perhaps lead to a misleading notion of centrality.

\begin{figure}[t]
    \centering
    \begin{subfigure}[b]{0.49\textwidth}
        \centering
        \includegraphics[width=\linewidth]{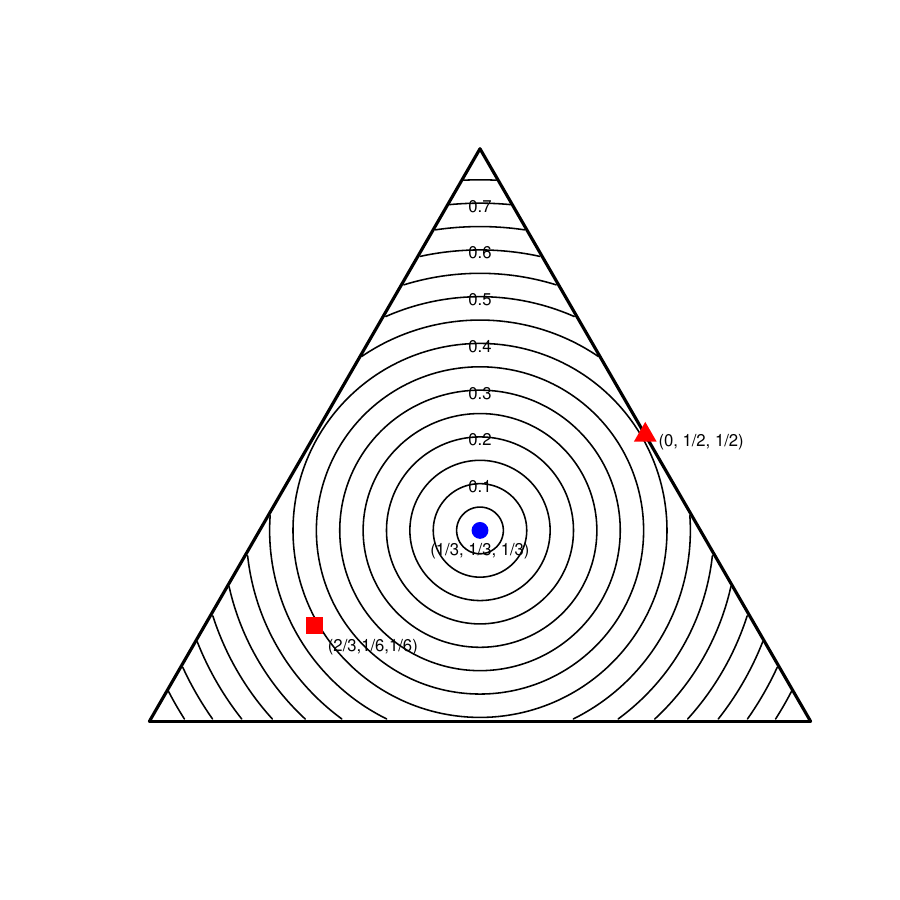}
        \caption{Euclidean metric}
        \label{fig: Euclid}
    \end{subfigure}
    \hfill
    \begin{subfigure}[b]{0.49\textwidth}
        \centering
        \includegraphics[width=\linewidth]{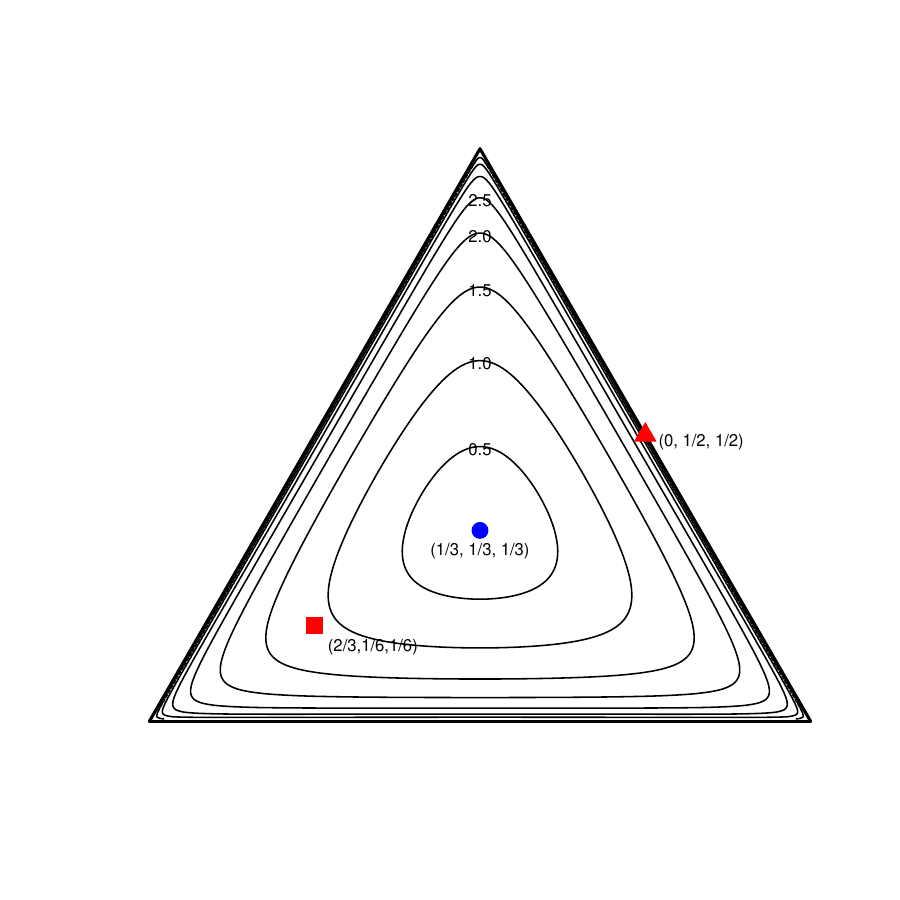}
        \caption{Aitchison metric}
        \label{fig: Aitchison}
    \end{subfigure}
    \caption{Comparison between the Euclidean metric and the Aitchison metric}
    \label{fig: comp-dist}
\end{figure}

To illustrate this issue, consider a three-part compositional outcome. Let $c_0=(1/3,1/3,1/3)$ denote the ``barycenter'' of the space of compositional data, and consider an interior point $c_1 = (2/3, 1/6, 1/6)$ and a boundary point $c_2 = (0, 1/2, 1/2)$. Under the Euclidean metric, as illustrated in Figure \ref{fig: Euclid}, the distances from the barycenter to these two points are identical: $d(c_0, c_1) = d(c_0, c_2)$.
From a compositional perspective, however, these two displacements are qualitatively different even if they have the same Euclidean magnitude.
The move from $c_0$ to $c_1$ remains within the interior of the simplex and corresponds to a finite reallocation of mass among components. In contrast, the move from $c_0$ to $c_2$ pushes the first component to zero, i.e., to the boundary (or an extreme point) of the simplex. This latter behavior can be considered more extreme. Accordingly, a metric that respects this kind of compositional geometry, such that $d(c_0, c_1) < d(c_0, c_2)$ as in Figure \ref{fig: Aitchison} based on the Aitchison metric, is more natural.

The same consideration applies in a more general setting considered in Section \ref{sec:gen}. Therefore, to extend Lee's (2009) bounding procedure to random object outcomes, we begin by introducing an appropriate metric that allows statistical analysis to reflect the geometry of the space in which outcomes take values. We then define an extended notion of the mean (i.e., \f mean) with respect to the metric and develop the Lee-type identified region for the \f mean of random object outcomes. Another key issue is how to operationalize the bounding approach. We address this issue by exploiting a well-behaved embedding that reduces our partial identification problem in a metric space to the one in a familiar Euclidean or Hilbert space.

\section{Benchmark example: Compositional data}\label{sec:bench}
Compositional data play a central role in economic analysis, as many key economic objects are inherently defined as allocations across mutually exclusive categories that sum to a whole. For example, household time allocation among labor, leisure, and housework is a fundamental concept for understanding labor supply decisions and intrahousehold bargaining. Analogously, the allocation of total expenditure across goods underpins classical consumer choice theory. In evaluating cash transfer programs, researchers likewise examine how households allocate consumption across categories; for example, the share devoted to health-related versus non-health food expenditures.

To accommodate these important scenarios, we now extend the scope of the Lee bounds to a compositional data scenario. Recall the treatment indicator ($D$) and selection indicators ($S_1, S_0, S$) defined as before. We retain the random assignment and monotonicity assumptions. Suppose that the potential outcomes $Y_0$ and $Y_1$ take values in the space for compositional data with positive components:
\begin{align*}
    \mathcal{Y} = \left\{y\in\mathbb{R}^k : y_j > 0\text{ for } j=1,\ldots,k,\text{ and } \, \sum_{j=1}^k y_j = 1\right\}.
\end{align*}
We impose the strict positivity condition $y_j > 0$ only to simplify the discussion that follows. Cases with zero components are accommodated within our general framework in Section~\ref{sec:gen}.

To introduce a notion of the expectation of the potential outcome $Y_t \in \mathcal{Y}$ for $t\in\{0,1\}$, let us consider the Aitchison metric $d$ on $\mathcal{Y}$ defined as
\begin{align*}
d(x,y) = \|\Psi(x) - \Psi(y)\|,
\end{align*}
where $\|\cdot\|$ is the Euclidean metric on $\mathbb{R}^k$ and 
\begin{align}
    \Psi(x) = \left(\log {\frac{x_1}  {g(x)}},\dots, \log {\frac{x_k} {g(x)}}\right)^{\prime},\quad g(x) = \left(\prod_{j=1}^k x_j\right)^{1/k}.\label{eq:Psi Aitchison}
\end{align}
The Aitchison metric is commonly applied for compositional data analysis (\citealp{Aitchison:1982}). As illustrated in Figure \ref{fig: Aitchison}, with the Aitchison metric, the resulting distances reflect the underlying geometric structure. Based on the metric space $(\mathcal{Y},d)$, we introduce the Fr\'echet mean of $Y_t\in\mathcal{Y}$, which is defined as
\begin{align*}
    \mathbb{E}_\oplus[Y_t]=\argmin_{y \in \mathcal{Y}}\E{d^2(y, Y_t)}.
\end{align*}
The Fr\'echet mean is a direct generalization of the conventional mean for the Euclidean space toward a general metric space, and its statistical analysis has been increasingly popular in recent literature (see, e.g., \citealp{Dubey:2024}). Analogously, the conditional Fr\'echet mean is defined as $\mathbb{E}_\oplus[Y_t \,| \,\cdot\,]=\mathrm{arg\,min}_{y \in \mathcal{Y}}\E{d^2(y, Y_t)\,|\,\cdot\,}$, and the average treatment effect for the always-observed group is characterized by some contrast of $\mu_{1\oplus}$ and $\mu_{0\oplus}$, where $\mu_{t\oplus}=\mathbb{E}_\oplus[Y_t\mid S_1=S_0=1]$.

As in the scalar outcome case, the assumptions of random assignment and monotonicity imply that $\mu_{0\oplus}$ is point identified as
\begin{equation}
    \mu_{0\oplus} = \mathbb{E}_\oplus[Y_0\mid S_1=S_0=1,D=0] = \mathbb{E}_\oplus[Y\mid S=1,D=0]. \label{eq:mu0}
\end{equation}
In contrast, $\mu_{1\oplus}$ is not point identified as in the scalar outcome case. Thus, our main task is to characterize the sharp identified set for $\mu_{1\oplus}$. To this end, we note that (i) the map $\Psi:\mathcal{Y} \to \mathbb{R}^k$ is an isometric and injective map from $\mathcal{Y}$ to $\mathbb{R}^k$, and (ii) the image space $\Psi(\mathcal{Y}) = \{x \in \mathbb{R}^k: \sum_{j=1}^k x_j =0\}$ is a closed convex subset of $\mathbb{R}^k$. Due to these properties, $\mu_{1\oplus}$ can be written as
\begin{align*}
    \mu_{1\oplus} = \Psi^{-1}\left(\E{\Psi(Y_1)\mid S_1=S_0=1}\right).
\end{align*}
Note that the expectation $\mathbb{E}[\cdot]$ on the right-hand side is the one for the conventional Euclidean vector, and this representation motivates the following two-step procedure to construct an identified set for $\mu_{1\oplus}$:
\begin{align*}
    \text{Step 1}&\text{: Derive the sharp identified set of $\E{\Psi(Y_1)\mid S_1=S_0=1}$ in $\mathbb{R}^k$ (denoted by $\mathcal{S}_1$),}\\
    \text{Step 2}&\text{: Obtain an identified set of $\mu_{1\oplus}$ in $\mathcal{Y}$ by taking the inverse image $\Psi^{-1}(\mathcal{S}_1)$.}
\end{align*}
For Step 1, a similar reasoning to the scalar-valued case applies, and the sharp identified set of $\E{\Psi(Y_1)\mid S_1=S_0=1}$ can be written as
\begin{align}
    \left\{
    \E{\Psi(Y)\mid D = 1, S = 1, \Psi(Y)\in \mathcal{A}}:
    \mathcal{A}\in\mathscr{A}
    \right\},\label{eq:B}
\end{align}
where $\mathscr{A}$ is the collection of all subsets $\mathcal{A}$ such that $\P{\Psi(Y)\in \mathcal{A} \mid  D = 1, S = 1} = p$ (the set in (\ref{eq:B}) coincides with the set $\mathcal{I}_1$ in our general result in Section \ref{sub:main}). Furthermore, a standard result in convex analysis implies that the identified set in (\ref{eq:B}) admits an alternative tractable characterization $\mathcal{S}_1$. For Step 2, we can show that $\Psi^{-1}(\mathcal{S}_1)$ is indeed the sharp identified set of $\mu_{1\oplus}$, so the main result of this section is summarized as follows.

\begin{proposition}
Consider the setup of this section (in particular, assume random assignment and monotonicity) with some regularity conditions (Assumption \ref{assumption:metric} below). Then, $\mu_{0\oplus}$ is point identified as in (\ref{eq:mu0}). Moreover, $\mu_{1\oplus}$ is partially identified and its sharp identified set is $\Psi^{-1}(\mathcal{S}_1)$. 
\end{proposition}

This proposition is obtained as a special case of the general result in the next section. Before turning to our general result, we illustrate the proposed strategy and its appeal using a numerical example.

\subsection{Numerical illustration}\label{sec:numerical}

This subsection illustrates our proposed procedure. We consider a hypothetical randomized controlled trial setting with sample attrition using data from the 2024 American Time Use Survey. We restrict attention to respondents aged 25-55 who are employed and report positive time spent on leisure, labor-market work, and housework.\footnote{We define leisure as the sum of first-tier codes 1 (Personal Care), 11 (Eating and Drinking), 12 (Socializing, Relaxing, and Leisure), and 13 (Sports, Exercise, and Recreation). Labor-market work corresponds to first-tier code 5 (Work \& Work-Related Activities). We define housework as the sum of first-tier codes 2 (Household Activities), 3 (Caring for \& Helping Household Members), 4 (Caring for \& Helping Nonhousehold Members), and 7 (Consumer Purchases). See \url{https://www.bls.gov/tus/lexicons/lexiconwex2024.pdf} for the list of detailed activities within each category.} 
The resulting sample size is $1,397$.
We then randomly assign individuals to two groups, treating the split as if it were a treatment-control assignment ($711$ in the control group and $686$ in the treatment group). 
To mimic attrition, we randomly drop observations according to group-specific selection probabilities. Specifically, we set $\P{S=1\mid D=1}=0.90$ and vary the control-group retention rate over $\P{S=1\mid D=0}\in\{0.85,0.70\}$. Since $\P{S=1\mid D=1}$ is held fixed, a smaller $\P{S=1\mid D=0}$ implies a smaller $p$, which in turn increases the degree of ``contamination'' $1-p$ and limits identifying power.
Because there is, in fact, no treatment, we expect no treatment effect and hence assess whether the estimated bounds $\hat{\mathcal{S}}_1$ contain the point-identified \f mean for the controlled group $\hat{\mu}_{0\oplus}$.

\begin{figure}[t]
    \centering
    \begin{subfigure}[b]{0.49\textwidth}
        \centering
        \includegraphics[width=\linewidth]{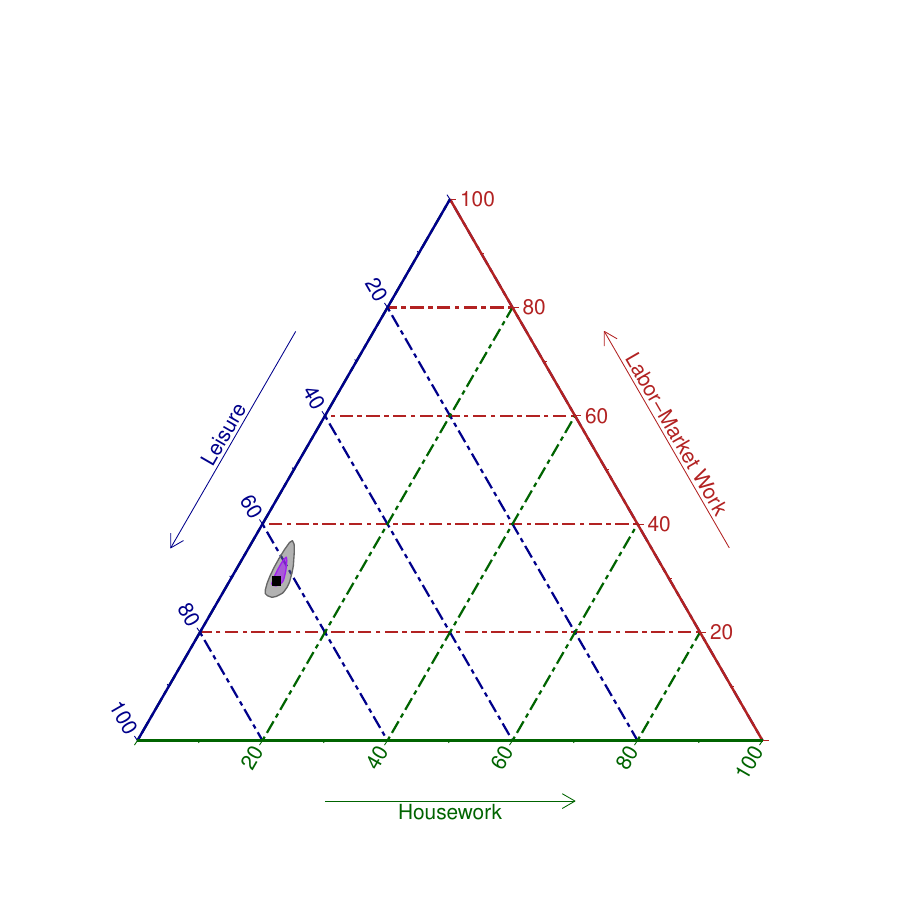}
        \caption{$\P{S=1\mid D=0}=0.85$}
        \label{fig: ex1}
    \end{subfigure}
    \hfill
    \begin{subfigure}[b]{0.49\textwidth}
        \centering
        \includegraphics[width=\linewidth]{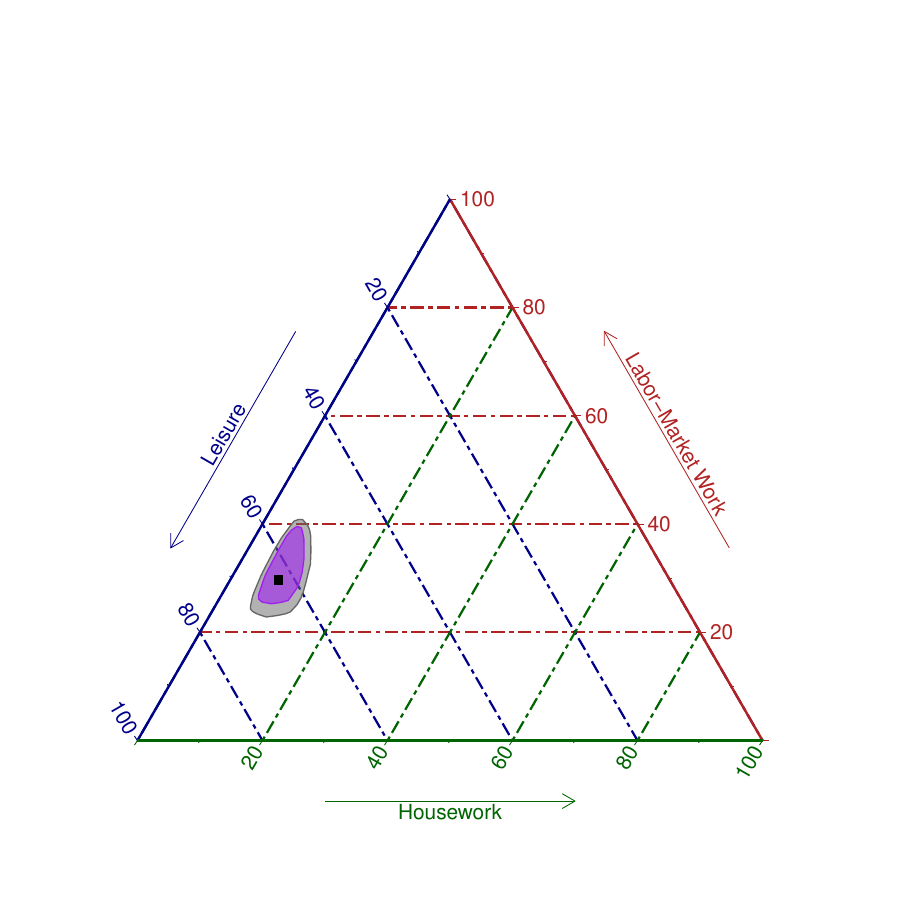}
        \caption{$\P{S=1\mid D=0}=0.70$}
        \label{fig: ex2}
    \end{subfigure}
    \caption{Ternary plots}
    \label{fig: ex}

    \begin{flushleft}
    \footnotesize
    \renewcommand{\baselineskip}{11pt}
    \textbf{Note:} The black square (\raisebox{0.15ex}{\scalebox{0.7}{$\blacksquare$}}) indicates the point estimate ${\hat{\mu}}_{0\oplus}$ of the point-identified mean ${\mu}_{0\oplus}$, the inner purple region corresponds to the estimated identified set ${\hat{\mathcal{S}}}_1$, and the outer gray region is the bootstrap confidence region for $\mathcal{S}_1$.
    \end{flushleft}
\end{figure}
Figure \ref{fig: ex} presents the estimated bounds $\hat{\mathcal{S}}_1$ (inner, purple region) and point-identified \f mean $\hat{\mu}_{0\oplus}$ (square dot), as well as the 95\% bootstrap confidence region for ${\mathcal{S}}_1$ (outer, gray region), whose validity is shown in Section \ref{sub:inf}. We can see that the bounds and confidence sets contain $\hat{\mu}_{0\oplus}$, suggesting the validity of our proposed bounds. In the low-attrition case (Figure \ref{fig: ex1}), the identified set is reasonably tight and informative. When the attrition rate is higher (Figure \ref{fig: ex2}), the identified set expands, but it still provides non-trivial identifying power.

\begin{table}[t]
    \centering
    \begin{tabular}{lcccc}\hline\hline
                          & $\hat{\mu}_{0\oplus}$ & Proj. $ \Psi^{-1}(\hat{\mathcal{S}}_1)$ & CI & Lee\\\hline
        \multicolumn{5}{l}{\textbf{A.} $\P{S=1\mid D=0}=0.85$}  \\                
        Leisure           & 0.630 & [0.593,     0.640] & [0.567,    0.660] & [0.553, 0.583]\\
        Labor market work & 0.295 & [0.286,     0.339] & [0.265,    0.369] & [0.313, 0.343]\\
        Housework         & 0.075 & [0.066,     0.086] & [0.058,    0.098] & [0.089, 0.109]\\\hline
        \multicolumn{5}{l}{\textbf{B.} $\P{S=1\mid D=0}=0.70$}  \\                
        Leisure           & 0.626 & [0.541,     0.676] & [0.529,    0.698] & [0.522, 0.612]\\
        Labor market work & 0.296 & [0.253,     0.396] & [0.228,    0.408] & [0.283, 0.388]\\
        Housework         & 0.077 & [0.049,     0.114] & [0.043,    0.130] & [0.064, 0.129]\\\hline
    \end{tabular}
    \caption{Bounds on $\mu_{1\oplus}$}
    \label{tab:empirical results}

    \begin{flushleft}
    \footnotesize
    \renewcommand{\baselineskip}{11pt}
    \textbf{Note:} The first column reports the point estimates of $\mu_{0\oplus}$, the second shows the estimated identified set $\Psi^{-1}(\mathcal{S}_1)$, projected onto each component of the composition; the third presents the corresponding confidence region, likewise projected onto each component; and the last column reports the conventional Lee bounds applied componentwise.
    \end{flushleft}
\end{table}
Table \ref{tab:empirical results} shows the projections of the estimated bounds and confidence region, as well as the naively applied standard Lee bounds. In Panel A, we can find that the componentwise Lee bounds do not contain the \f mean for the control group (which coincides with the counterfactual \f mean as well in our scenario) in any dimensions. This pattern suggests that the standard Lee bounds may fail to deliver a valid identified set for the center of the counterfactual outcome distribution (as measured by the \f mean).
By contrast, the componentwise projections of our proposed bounds do cover $\hat{\mu}_{0\oplus}$. 
We also note that these confidence intervals are simultaneously valid, as they are obtained by projecting the confidence region for $\Psi^{-1}(\mathcal{S}_1)$.

In Panel B, although the naive Lee bounds become wider, the bounds for the leisure component still fail to cover $\hat{\mu}_{0\oplus}$. In contrast, the componentwise projections of our proposed bounds remain valid, although they are less informative than in the low-attrition case shown in Panel A.

\section{Generalization: Lee bounds for random objects}\label{sec:gen}
\subsection{Setup and examples}

This section extends the benchmark example in the last section and presents a general framework to conduct partial identification analysis for the sample selection model in (\ref{eq:obs}), where the outcome variable $Y$ resides in a metric space $(\mathcal{Y},d)$.\footnote{Although we focus on the sample selection model, our analysis also applies to the contaminated data model studied by \cite{Horowitz_Manski:1995}; see Remark \ref{rem:HM} below.} To begin with, we redefine the notation and formally introduce our setup.

\begin{assumption}\label{assumption:setup}
Let $Y_0, Y_1 \in \mathcal{Y}$ be potential outcomes and $S_0, S_1 \in \{0,1\}$ be potential selection indicators for a treatment $D \in \{0,1\}$. $\mathcal{Y}$ is equipped with a metric $d$. Assume that $D$ is independent from $(Y_1, Y_0, S_1, S_0)$ (random assignment) and $S_1 \geq S_0$ with probability one (monotonicity). We observe $D$ and $(S, Y)$ generated from the selection model in (\ref{eq:obs}).
\end{assumption}
This setup is identical to the one in \cite{Lee:2009} except that the support $\mathcal{Y}$ can be non-Euclidean, and we are concerned with causal inference by a randomized experiment in the presence of sample selection. The monotonicity assumption $S_1 \ge S_0$, which rules out ``defiers'' ($S_1=0, S_0=1$) from the population, requires that the assignment $D$ can affect the sample selection only in one direction. This assumption is standard in many econometric models and aligns well with the threshold-crossing model \citep{Vytlacil:2002}.

In this setup, the average treatment effect of interest is characterized by some contrast of the Fr\'echet means 
\[
\mu_{t\oplus}=\mathbb{E}_\oplus[Y_t\mid S_1=S_0=1]=\argmin_{y \in \mathcal{Y}}\E{d^2(y,Y_{t})\mid S_{1}=S_{0}=1},
\]
for $t \in \{0,1\}$. Again, $\mu_{0\oplus}$ is point identified as in (\ref{eq:mu0}) under Assumption \ref{assumption:setup}. Thus, we focus on partial identification of $\mu_{1\oplus}$, and impose the following assumptions on the metric space $(\mathcal{Y},d)$.

\begin{assumption}\label{assumption:metric}
\quad
\begin{description}
    \item[(i)] There exists an injective map $\Psi$ from $\mathcal{Y}$ to a separable Hilbert space $\mathcal{H}$ with the metric $d_\mathcal{H}$ such that $d(x,y)=d_\mathcal{H}(\Psi(x),\Psi(y))$. Furthermore, $\Psi(\mathcal{Y})$ is a closed convex subset of $\mathcal{H}$. 
    \item[(ii)] $\E{\| \Psi(Y) \|^2} < \infty$, and the inner product $\langle u, \Psi(Y)\rangle$ induced by $d_\mathcal{H}$ is continuously distributed conditional on $D = 1$ and $S = 1$ for each $u\in\mathcal{H}$ with $\|u\| = 1$.
\end{description}
\end{assumption}

Assumption \ref{assumption:metric} (ii) is a regularity condition requiring the existence of the second moment and continuity of the distribution. The key assumption in our analysis is hence Assumption \ref{assumption:metric} (i). Assumption \ref{assumption:metric} (i) enables us to characterize the (conditional) Fr\'echet mean of the random variable $Y$ using the map $\Psi$. In fact, the (conditional) Fr\'echet mean of $Y$ is obtained by pulling back, via $\Psi^{-1}$, the (conditional) expectation of the embedded variable $\Psi(Y)$ in the Hilbert space $\mathcal{H}$, which is uniquely determined by the Riesz representation theorem. We refer to \cite{kuri:25} for further details on this point. Many of the common metric spaces studied in the literature of metric statistics are known to satisfy Assumption \ref{assumption:metric} (i). Here, we list some examples of metric spaces that are commonly used in statistical analysis of metric space-valued data. Additional examples are contained in Appendix \ref{app:ex}.

\begin{example}[Interval data] \label{ex:int}
    Set-valued data have a wide range of applications, for example, in census and survey data analysis and in biostatistics \citep{BeMo08,adusumilli2017empirical,li2021local}. For $d\geq 1$, let $K_{kc}(\mathbb{R}^d)$ denote the collection of nonempty compact and convex subsets of $\mathbb{R}^d$ and let $K_{kc}^B(\mathbb{R}^d)=\{F \in K_{kc}(\mathbb{R}^d): \sup_{f \in F}\|f\|\leq B\}$ where $\|\cdot\|$ is the standard Euclidean metric and $B$ is a positive constant. We can introduce a metric $d_{kc}$ on $K_{kc}(\mathbb{R}^d)$ defined as $d_{kc}(F,G)=\sqrt{\int_{\mathbb{S}^{d-1}}\{s_F(t)-s_G(t)\}^2\,dt}$, where $s_F(t)=\sup_{f \in F}t'f$ is the support function of $F$ and $\mathbb{S}^{d-1}=\{x \in \mathbb{R}^d:\|x\|=1\}$ is the $(d-1)$-dimensional unit sphere. In this case, the mapping $\Psi(F) = s_F(\cdot)$ is an isometry from $K_{kc}^B(\mathbb{R}^d)$ to $L^2(\mathbb{S}^{d-1})$, the space of functions such that $\int_{\mathbb{S}^{d-1}}f^2(x)dx<\infty$, and the image space $\Psi(K_{kc}^B(\mathbb{R}^d))$ is a closed convex subset of $L^2(\mathbb{S}^{d-1})$ \citep[Lemma 2.1]{kuri:25}. In particular, for $d=1$, $K_{kc}(\mathbb{R})=\{[a,b]: -\infty<a\leq b<\infty\}$ is the space for interval data, and its image space $\Psi(K_{kc}(\mathbb{R}))$ can be identified with $\{(x,y) \in \mathbb{R}^2: y+x\geq 0\}$, which is a closed convex subset of $\mathbb{R}^2$.
\end{example}

\begin{example}[Functional data]\label{ex:fun}
    Data consisting of functions are referred to as functional data. Such data are prevalent in longitudinal studies, environmental monitoring, and biomedical imaging \citep{ramsay2005functional, hsing2015theoretical, wang2016functional}. Those functions are typically assumed to be situated in the space $L^2(\mathcal{I})$ of real-valued square-integrable functions on a compact interval $\mathcal{I}$. This space carries the inner product $\langle f_1, f_2 \rangle_{L^2}=\int_{\mathcal{I}} f_1(x)f_2(x)\,dx$ and the corresponding $L^2$ metric, given by $\|f_1-f_2\|_{L^2} = \sqrt{\int_\mathcal{I}\{f_1(x) - f_2(x)\}^2\,dx}$. The space $L^2(\mathcal{I})$ is closed and convex with respect to the $L^2$ metric. In this example, we can set $\mathcal{H} = L^2(\mathcal{I})$ and take $\Psi$ as the identity map $\Psi(f) = f$. 
\end{example}

\begin{example}[One-dimensional probability distributions]\label{ex:dist}
    Distributional data arise when each data point is regarded as a probability distribution, and they have been applied in many fields, including economics and multi-cohort studies \citep{petersen2022modeling,Gunsilius:2023,zhou2024wasserstein}. Let $\mathcal{W}$ denote the space of Borel probability measures on a real line, with finite second moments. This space becomes a metric space when equipped with the 2-Wasserstein distance $d_\mathcal{W}(\mu_1, \mu_2) = \sqrt{\int_0^1 \{F_{\mu_1}^{-1}(u) - F_{\mu_2}^{-1}(u)\}^2\,du}$. Here, $F_{\mu}^{-1}$ denotes the quantile function of a probability distribution $\mu$. The metric space $(\mathcal{W}, d_\mathcal{W})$ is called the Wasserstein space \citep{panaretos2020invitation}, and distributional data are typically assumed to reside within this space. The mapping $\Psi(\mu) = F_{\mu}^{-1}$ is clearly an isometry from $\mathcal{W}$ to the Hilbert space $L^2([0, 1])$. Moreover, the image $\Psi(\mathcal{W})$ is characterized as the set of all square-integrable, almost everywhere increasing functions on $[0, 1]$, and it is closed and convex in $L^2([0, 1])$ \citep[Proposition 2.1,][]{bigot2017geodesic}.
\end{example}

\begin{example}[Networks]
    Consider a simple, undirected, and weighted network with a set of nodes $\{v_1, \ldots, v_m\}$ and a set of bounded edge weights $ \{w_{pq}: p, q=1, \ldots, m\}$, where  $0 \le w_{pq} \le W$. Such a network can be uniquely represented by its graph Laplacian matrix $L = (l_{pq}) \in \mathbb{R}^{m^2}$, defined as 
    \[
    l_{pq} = 
    \begin{cases}
        -w_{pq} & \text{if}\,\, p \neq q, \\
        \sum_{r \neq p}w_{pr} & \text{if}\,\, p = q, 
    \end{cases}
     \quad 
        p, q = 1, \ldots, m.
    \]
    The space of graph Laplacians is given by $\mathcal{L}_m = \{L = (l_{pq}): L = L', L1_m = 0_m, -W \le l_{pq} \le 0 \,\, \text{for}\,\, p \neq q\}$ where $1_m$ and $0_m$ are the $m$-vectors of ones and zeros, respectively. This space provides a natural framework for characterizing network structures such as social connections, transportation links, or gene regulator networks \citep{kola:14, zhou2022network,severn2022manifold}. Equipped with the Frobenius metric $d_F$ defined as $d_F(L_1,L_2) = [\text{tr}\{(L_1-L_2)'(L_1-L_2)\}]^{1/2}$, the space of graph Laplacians $\mathcal{L}_m$ forms a closed and convex subset of the Euclidean space $\mathbb{R}^{m^2}$  \citep[Proposition 1]{zhou2022network}.
\end{example}

\subsection{Main results}\label{sub:main}
Now, we are in a position to derive the sharp identified set of $\mu_{1\oplus}=\mathbb{E}_\oplus[Y_1\mid S_1=S_0=1]$ (see Definition \ref{def} for the formal definition of the sharp identified set). Under Assumption \ref{assumption:metric}, the potential Fr\'echet mean can be expressed as $\mu_{1\oplus}=\Psi^{-1}(\E{\Psi(Y_{1})|S_{1}=S_{0}=1})$. Thus, as in the last section, we begin by characterizing an identified set for $\E{\Psi(Y_{1})|S_{1}=S_{0}=1}$. Let $p= \P{S=1 | D=0}/\P{S=1 | D=1}$ and observe that
\[
\P{\Psi(Y)\in\mathcal{A}|D = 1, S = 1} = p\P{\Psi(Y_{1})\in\mathcal{A}|S_{1} = S_{0} = 1} + (1-p)\P{\Psi(Y_{1})\in\mathcal{A}|S_{1} = 1, S_{0} = 0},
\]
where $\mathcal{A}$ is an arbitrary measurable subset of $\mathcal{H}$. 
Hence, a $100\times p\%$ of the observed subpopulation with $D = 1, S = 1$ corresponds to the always-observed group. 
Then, by letting $\mu(\mathcal{A})\coloneqq \P{\Psi(Y)\in\mathcal{A}|D = 1, S = 1}$, the sharp identified set for $\E{\Psi(Y_1) \mid S_1=S_0=1}$ is obtained as
\begin{align}
    \mathcal{I}_1 \coloneqq \left\{\frac{1}{p} \int_{\mathcal{H}} x\,d\nu(x) : 0\leq \nu(\mathcal{A}^\prime)\leq \mu(\mathcal{A}^\prime)\text{ for any measurable } \mathcal{A}^\prime\subset\mathcal{H},\text{ and } \nu(\mathcal{H})=p\right\},\label{eq:sharpset}
\end{align}
where $\int_{\mathcal{H}} \cdot \,d\nu$ means the Bochner integral, and the sharpness of $\mathcal{I}_1$ is shown in Appendix \ref{sec:sharp}.
Our partial identification results are presented as follows.
\begin{theorem}\label{thm}
Suppose Assumptions \ref{assumption:setup} and \ref{assumption:metric} hold true. Then
\begin{description}
    \item[(i)] The convex closure of $\mathcal{I}_1$ satisfies
    \begin{equation}
    \overline{\mathrm{conv}}\,\mathcal{I}_1 = 
    \bigcap_{\substack{u\in\mathcal{H}\\ \|u\|_\mathcal{H}=1}}\left\{v\in\mathcal{H} : \langle u, v \rangle \leq \sigma_{\mathcal{I}_1}(u)\right\} \eqqcolon \mathcal{S}_1, \label{eq:S1}
    \end{equation}
    where 
    \[
    \sigma_{\mathcal{I}_1}(u)\coloneqq \E{\langle u, \Psi(Y)\rangle\mid  D = 1, S = 1, \langle u, \Psi(Y)\rangle \geq Q_{\langle u, \Psi(Y)\rangle}(1-p)},
    \]
    and $Q_{\langle u, \Psi(Y)\rangle}(1-p)$ denotes the $(1-p)$-th conditional quantile of $\langle u, \Psi(Y)\rangle$ given $D = 1$ and $S = 1$. In other words, $\mathcal{S}_1$ is a valid identified set of $\E{\Psi(Y_{1})|S_1=S_0=1}$.
    \item[(ii)] $\Psi^{-1}(\mathcal{S}_1)$ is a valid identified set of $\mu_{1\oplus}$.
    \item[(iii)] When $\Psi(Y)$ is finite dimensional, it holds $\mathcal{I}_1=\mathcal{S}_1$, i.e., $\mathcal{S}_1$ is the sharp identified set of $\E{\Psi(Y_{1})|S_1=S_0=1}$.
    Furthermore, $\Psi^{-1}(\mathcal{S}_1)$ is the sharp identified set of $\mu_{1\oplus}$.
\end{description}
\end{theorem}

Theorem \ref{thm} establishes a Lee-type identified set for the conditional \f mean $\mu_{1\oplus}$ of general random objects. Theorem \ref{thm} (iii) shows that, when $\Psi(Y)$ is finite-dimensional, the proposed procedure delivers a sharp identified set for $\mu_{1\oplus}$. Representing a closed convex set as in \eqref{eq:S1} is standard in econometrics (e.g., \citealp{BeMo08}), and this representation is computationally feasible in practice.
Further details are discussed in Section \ref{sec:implementation}. We note that incorporating covariates is straightforward (see Remark \ref{rem:cov} below).

When $\Psi(Y)$ is infinite-dimensional, evaluating \eqref{eq:S1} may be infeasible, and $\mathcal{S}_1$ need not coincide with the sharp identified set for $\mu_{1\oplus}$. Nevertheless, one may still be able to obtain a sharp identified set over a finite grid of points by applying the same procedure to a finite-dimensional projection of $\Psi(Y)$, which is useful to approximate $\mathcal{S}_1$ or to obtain a feasible, valid identified set. See Remark \ref{rem:feasible} below for the case of distributional outcomes.

\begin{remark}[Covariates]\label{rem:cov}
    Suppose that a baseline covariates vector $X$ with support $\mathcal{X}\subset\mathbb{R}^k$ is available.
    The previous argument is valid conditional on $X=x$ so that we can obtain the identified set $\mathcal{S}_1(x)$ for each $x\in\mathcal{X}$.
    Then the identified set of $\E{\Psi(Y_{1})|S_1=S_0=1}$ is given by $\mathcal{S}_1^X= \{\E{s(X)\mid S=1,D=0}: s(\cdot)\text{ is measurable and }s(x) \in \mathcal{S}_1(x),\,\forall x\in\mathcal{X}\}$, or equivalently, 
    \begin{align*}
        \mathcal{S}_1^X = \bigcap_{\substack{u\in\mathcal{H}\\ \|u\|_\mathcal{H}=1}}\left\{v\in\mathcal{H} : \langle u, v \rangle \leq \E{\sigma_X(u) \mid S=1,D=0}\right\},
    \end{align*}
    where $\sigma_x(u)$ is the conditional analog of $\sigma_{\mathcal{I}_1}(u)$. The identified set of $\mu_{1\oplus}$ can be obtained by $\Psi^{-1}(\mathcal{S}_1^X)$. When the covariates are discrete, or when continuous covariates are discretized as in \cite{Lee:2009}, the identified set is simply given by
    \begin{align*}
        \mathcal{S}_1^X = \bigcap_{\substack{u\in\mathcal{H}\\ \|u\|_\mathcal{H}=1}}\left\{v\in\mathcal{H} : \langle u, v \rangle \leq \sum_{x\in\mathcal{X}} w_x\sigma_{x}(u)\right\},
    \end{align*}
    where $w_x = \P{X=x \mid S=1, D=0}$. \hfill$\square$
\end{remark}

\begin{remark}[Distributional outcome]\label{rem:feasible}
    Recall the one-dimensional distributional outcome case discussed in Example \ref{ex:dist} (i.e., $\mathcal{Y} = \mathcal{W}$). Then $\Psi(\mathcal{Y})$ can be regarded as the space of quantile functions. Take a finite number of $0<q_1<\cdots<q_k<1$ ``evaluation points'' and consider a map $\Pi : \mathcal{Y} \ni Y \mapsto (F_Y^{-1}(q_1),\ldots, F_Y^{-1}(q_k)) \in \mathbb{R}^k$. Note that the image space $\Pi(\mathcal{Y})=\{(x_1,\dots,x_k) \in \mathbb{R}^k: -\infty<x_1\leq \dots \leq x_k<\infty\}$ is a closed convex subset of $\mathbb{R}^k$. As a result, by applying Theorem \ref{thm} (iii), we can derive the sharp bounds on $\E{\Pi(Y_{1})|S_1=S_0=1}$ by
    \begin{align*}
        \mathcal{S}_{1,\Pi}= \bigcap_{\substack{u\in\mathbb{R}^k\\ \|u\|=1}}\left\{v\in\mathbb{R}^k : \langle u, v \rangle \leq \sigma(u)\right\}, 
    \end{align*}
    where $\sigma(u)= \mathbb{E}[\langle u, \Pi(Y)\rangle\mid  D = 1, S = 1, \langle u, \Pi(Y)\rangle \geq Q_{\langle u, \Pi(Y)\rangle}(1-p)]$.
    Then, we can construct the following valid identified set of $\E{\Psi(Y_{1})|S_1=S_0=1}$: 
    \begin{align*}
        \mathcal{I}_1 \subset \left\{f\in\Psi(\mathcal{W}) : \left(f(q_1),\ldots,f(q_k)\right) \in \mathcal{S}_{1,\Pi}\right\}.
    \end{align*}  
    In practice, although the right-hand side is still infeasible, we can approximate and visualize this identified set by taking a large $k$, picking a fine grid of points in $\mathcal{S}_{1,\Pi}$, and then drawing the linearly interpolated values $f(q_1),\ldots,f(q_k)$. 
    Alternatively, one can rely on a feasible superset $[L(\cdot), U(\cdot)]$, where $L(q)\coloneqq s_j^\ell$ if $q\in[q_j,q_{j+1})$ and $U(q)\coloneqq s_{j+1}^u$ if $q\in(q_j,q_{j+1}]$, and $[s_j^\ell, s_j^u]$ denotes the projection of $\mathcal{S}_{1,\Pi}$ onto the $j$-th element.\hfill$\square$
\end{remark}

We conclude our identification analysis with a remark on the contaminated and corrupted data models in the sense of \citet{Horowitz_Manski:1995}.
\begin{remark}[Contaminated and corrupted data model]\label{rem:HM}
    We begin with the standard contaminated data model. Suppose we are interested in the mean of $Y_1 \in \mathbb{R}$, but only observe the contaminated sample $Y = Y_1 Z + Y_0 (1-Z) \in \mathbb{R}$, where $Z\in\{0,1\}$ is the (non-)contamination indicator taking zero if the realization is contaminated. Suppose the contamination $Z$ is independent of our interest $Y_1$.
    \citet[Corollary 4.1]{Horowitz_Manski:1995} provides the sharp bounds on $\E{Y_1}$ under the assumption that $\P{Z=0}\leq \lambda <1$, where $\lambda$ is a known or identified constant.
    Our analysis in this section, for the case where $Y,Y_1,Y_0\in\mathcal{Y}$, immediately applies to this context.
    In particular, we have that $\mathbb{E}_{\oplus}[Y_1] \in \Psi^{-1}(\mathcal{S})$, where 
    \begin{equation*}
        \mathcal{S} = 
        \bigcap_{\substack{u\in\mathcal{H}\\ \|u\|_\mathcal{H}=1}}\left\{v\in\mathcal{H} : \langle u, v \rangle \leq \sigma(u)\right\},
    \end{equation*}
    with $\sigma(u)= \E{\langle u, \Psi(Y)\rangle\mid  \langle u, \Psi(Y)\rangle \geq Q_{\langle u, \Psi(Y)\rangle}(\lambda)}$, and $Q_{\langle u, \Psi(Y)\rangle}(q)$ denotes the $q$-th quantile of $\langle u, \Psi(Y)\rangle$.

    Suppose that $Z$ may be correlated with $Y_1$ (i.e., corrupted data case) but we know that $Y_1\in\mathcal{Y}_1 \subset \mathcal{Y}$ with known $\mathcal{Y}_1$ such that $\Psi(\mathcal{Y}_1)$ is closed and convex.
    The previous argument suggests that $\E{\Psi(Y_1)\mid Z=1} \in \mathcal{S}$, and hence we obtain the identified set
    \begin{align*}
        \E{\Psi(Y_1)} \in (1-\P{Z=0}) \mathcal{S} \oplus \P{Z=0} \Psi(\mathcal{Y}_1)
        \subset (1-\lambda) \mathcal{S} \oplus \lambda \Psi(\mathcal{Y}_1),
    \end{align*}
    where the last inclusion uses the convexity of $\Psi(\mathcal{Y}_1)$ and $\mathcal{S}\subset\Psi(\mathcal{Y}_1)$. \hfill$\square$
\end{remark}

\subsection{Characterizing treatment effects}\label{sec:treatment effect}
Once the sharp identified set for $\mu_{1\oplus}$ is obtained, the remaining question is how to summarize the treatment effect. Unlike the Euclidean-outcome case, when outcomes take values in a general metric space $\mathcal{Y}$, one cannot in general define treatment effects by algebraic differencing of potential means, since $\mathcal{Y}$ need not admit a linear structure. Nevertheless, researchers can define treatment effects in ways that are tailored to their substantive objectives. Indeed, there can be several approaches to defining causal effects, which we list below.

\textbf{Difference in the embedded space.}
The first possibility is to define the causal effect through differences in an embedded space. For a general metric space that can be embedded into a Hilbert space $\mathcal{H}$, we can define the contrast $\Psi(\mu_{1\oplus})-\Psi(\mu_{0\oplus})$ as the causal effect of interest, provided that the embedded object $\Psi(Y)$ has a reasonable interpretation.

A leading example is the case in which $\mathcal{Y}$ is the $2$-Wasserstein space for distributional outcomes. In this setting, it is common to work with quantile-function representations and to define a quantile-treatment-effect-type parameter as the difference in the embedded space of quantile functions $\Psi(\mathcal{Y})$ (e.g., \citealp{LiKoWa23} and \citealp{va:25}):
\begin{align*}
    \E{\Psi(Y_1)|S_1=S_0=1}-\E{\Psi(Y_0)|S_1=S_0=1} = F_{\mu_{1\oplus}}^{-1}(\cdot)-F_{\mu_{0\oplus}}^{-1}(\cdot).
\end{align*}
In this case, the identified set for the embedded contrast can be written as the Minkowski difference
\begin{align*}
    \mathcal{S}_1 \ominus \{\Psi(\mu_{0\oplus})\} =
    \left\{\nu - \E{\Psi(Y_0)|S_1=S_0=1} : \nu \in \mathcal{S}_1\right\}.
\end{align*}

\textbf{Difference in the projections.}
A second option is to focus on a real-valued contrast obtained by projecting the mean objects $\mu_{0\oplus}$ and $\mu_{1\oplus}$ onto $\mathbb{R}$.
For example, suppose $Y$ is a compositional outcome and the researcher's interest is in the percentage-point change in the first component. 
Let $\Pi:\mathcal{Y}\to\mathbb{R}$ denote the projection onto the first coordinate, i.e., $\Pi(y)=y_1$ for $y=(y_1,\ldots,y_k)\in\mathcal{Y}$. 
Then a natural causal estimand is $\Pi(\mu_{1\oplus})-\Pi(\mu_{0\oplus})$. 
More generally, the same construction applies whenever $\Pi:\mathcal{Y}\to\mathbb{R}$ has a meaningful interpretation in the application at hand.
In this case, an identified set for the projected contrast is given by
\begin{align*}
    \Pi\left(\Psi^{-1}(\mathcal{S}_1)\right) \ominus \Pi\left(\mu_{0\oplus}\right)
    = \left\{\nu - \Pi\left(\mu_{0\oplus}\right) : \nu \in \Pi\left(\Psi^{-1}(\mathcal{S}_1)\right)\right\}.
\end{align*}

\textbf{Geodesics.}
As a third option, one may employ the geodesic average treatment effect (GATE) proposed by \citet{kurisu_etal:24}. 
The GATE is defined as the entire geodesic path from $\mu_{0\oplus}$ to $\mu_{1\oplus}$:
\begin{align*}
    \gamma_{\mu_{0\oplus},\mu_{1\oplus}} & = 
    \left\{\gamma_{\mu_{0\oplus},\mu_{1\oplus}}(t): t \in [0,1]\right\} \\ 
    & = \left\{\Psi^{-1}((1-t)\Psi(\mu_{0\oplus}) + t\Psi(\mu_{1\oplus})): t \in [0,1]\right\}. 
\end{align*}
Intuitively, for $\alpha, \beta\in\mathcal{Y}$, the geodesic $\gamma_{\alpha,\beta}$ is defined as the shortest path from $\alpha$ to $\beta$. For a precise definition of geodesics in a general metric space and for the relationship between GATE and the conventional ATE defined for Euclidean outcomes, see \cite{kurisu_etal:24}. 
The full geodesic includes both the direction and the magnitude of the change induced by the treatment. This provides insight into how the underlying object evolves along the shortest intrinsic path.
Based on the identified set $\Psi^{-1}(\mathcal{S}_1)$ for $\mu_{1\oplus}$, the identified set of the GATE $\gamma_{\mu_{0\oplus},\mu_{1\oplus}}$ is given by the following set of geodesics:
\[
\left\{ \gamma_{\mu_{0\oplus},\mu} : \mu\in\Psi^{-1}(\mathcal{S}_1) \right\}.
\]
Provided that $\mathcal{S}_1$ is sharp, the induced identified sets for these causal estimands are sharp as well.

\section{Implementation}\label{sec:implementation}
This section provides detailed estimation and inference procedures for finite-dimensional outcomes. In the infinite-dimensional case, the same approach applies after projecting the object of interest onto a finite-dimensional parameter.

\subsection{Estimation}
We begin with the estimation procedure.
Recalling that
\begin{align*}
    \mu_{0\oplus} = \mathbb{E}_\oplus[Y|S=1,D=0] = 
    \Psi^{-1}\left(\E{\Psi(Y) \mid S=1, D=0}\right),
\end{align*}
the \f mean $\mu_{0\oplus}$ can be estimated by its sample analog, or $\Psi^{-1}(\hat{\mathbb{E}}[\Psi(Y) \mid S=1, D=0])$, where $\hat{\mathbb{E}}[\Psi(Y) \mid S=1, D=0])$ is the standard sample conditional mean in the Euclidean space.
The sharp identified set of $\mu_{1\oplus}$ can be estimated by
\begin{align}
    \hat{\mathcal{S}}_1 = 
    \bigcap_{u\in \mathbb{S}^{d-1}}\left\{v\in\mathbb{R}^d : \langle u, v \rangle \leq \hat{\sigma}_{\mathcal{I}_1}(u)\right\},\label{eq:Shat}
\end{align}
where $\mathbb{S}^{d-1}=\{x \in \mathbb{R}^d:\|x\|=1\}$ is the $(d-1)$-dimensional unit sphere,
\begin{align*}
    \hat{\sigma}_{\mathcal{I}_1}(u)= 
    \frac{\sum_{i=1}^{n} \langle u, \Psi(Y_i)\rangle S_i D_i \mathbf{1}\left\{\langle u, \Psi(Y_i)\rangle \geq \hat{Q}_{\langle u, \Psi(Y_i)\rangle}(1-\hat{p})\right\} }{\sum_{i=1}^{n} S_i D_i \mathbf{1}\left\{\langle u, \Psi(Y_i)\rangle \geq \hat{Q}_{\langle u, \Psi(Y_i)\rangle}(1-\hat{p})\right\}},
\end{align*}
$\hat{p} = \{(\sum_{i=1}^n S_i(1-D_i))/(\sum_{i=1}^n (1-D_i))\}/\{(\sum_{i=1}^n S_i D_i)/(\sum_{i=1}^n D_i)\}$, and $\hat{Q}_{\langle u, \Psi(Y)\rangle}(q)$ denotes the $q$-th sample conditional quantile of $\langle u, \Psi(Y)\rangle$ given $D = 1, S = 1$.
In practice, the intersection ``$\bigcap_{u\in \mathbb{S}^{d-1}}$'' in \eqref{eq:Shat} can be approximated arbitrarily well by evaluating \eqref{eq:Shat} on a sufficiently fine grid of points on the unit sphere $\mathbb{S}^{d-1}$. 
When $d$ is small, one may use a grid with (approximately) equal angular spacing. For example, see \cite{BeMoMo10} for the case $d=2$; for $d=3$, one may employ a Fibonacci lattice. 
In higher-dimensional cases, an approximation based on random number generation is often more convenient. For instance, if $Z\sim \mathcal{N}(\mathbf{0}, \mathbf{I}_d)$, then $Z/\|Z\|$ is uniformly distributed on the $(d-1)$-dimensional unit sphere \citep[Theorem 4.1]{Devroye:1986}, and thus can be used to form an approximation of the intersection.

\subsection{Inference}\label{sub:inf}
This subsection discusses a method for uncertainty quantification. 
For $\mu_{0\oplus}$, the quantity $\E{\Psi(Y)\mid S=1,D=0}$ is point identified and corresponds to a standard Euclidean conditional mean. Accordingly, conventional procedures (e.g., the bootstrap) apply, and one can construct a confidence region for $\E{\Psi(Y)\mid S=1,D=0}$, denoted by $\hat{\mathcal{R}}_0$, in a standard manner. The corresponding confidence region for $\mu_{0\oplus}$ is then given by $\Psi^{-1}(\hat{\mathcal{R}}_0)$.

We therefore focus on uncertainty quantification for the sharp identified set $\mathcal{S}_1$.
Let $\alpha\in(0,1)$ be a researcher-specified significance level. For each $b\in\{1,\ldots, B\}$, compute 
\begin{align}
    T(b) = \sup_{u\in\mathbb{S}^{d-1}}
    \frac{\tilde{\sigma}_{\mathcal{I}_1}^{(b)}(u) - \hat{\sigma}_{\mathcal{I}_1}(u)}{\sqrt{\hat{V}_{\sigma}(u)/n}},\label{eq:stat}
\end{align}
where $\tilde{\sigma}_{\mathcal{I}_1}^{(b)}(u)$ is the bootstrap counterpart of $\hat{\sigma}_{\mathcal{I}_1}(u)$, and $\hat{V}_{\sigma}(u)$ is the estimator of the asymptotic variance $V_{\sigma}(u)$ of $\hat{\sigma}_{\mathcal{I}_1}(u)$, which is given by \citet[Proposition 3, $V^{\mathrm{UB}} + V_C$ in his notation]{Lee:2009}. In practice, this supremum is approximated by the maximum over some fine grid points. Then, we compute the critical value
\begin{align*}
    \hat{\text{c.v.}} = 
    \text{the } (1-\alpha)\text{-th quantile of } \{T(b) : b=1,\ldots,B\},
\end{align*}
and obtain the confidence region of $\mathcal{S}_1$ as
\begin{align*}
    \hat{\mathcal{R}}_1= \bigcap_{u\in \mathbb{S}^{d-1}}\left\{v\in\mathbb{R}^d : \langle u, v \rangle \leq \hat{\sigma}_{\mathcal{I}_1}(u) + \hat{\text{c.v.}} \sqrt{\hat{V}_{\sigma}(u)/n}\right\},
\end{align*}
where $\mathbb{S}^{d-1}$ is the $(d-1)$-dimensional unit sphere.
The following proposition establishes the validity of this procedure.

\begin{assumption}\label{assumption:bootstrap}
\quad
    \begin{description}
        \item[(i)] $\Psi(Y)\mid (S=1,D=1)$ has a bounded, closed, and convex support $K\subset\mathbb{R}^d$.
        Moreover, $\Psi(Y)$ admits a conditional density $f_{\Psi(Y)}$ given $(S=1,D=1)$ that is continuous on $K$ and satisfies $0<\underline{f} \leq f_{\Psi(Y)}(x) \leq \overline{f} <\infty$ for all $x\in K$.
        \item[(ii)] $0<\E{SD}$, $0<\E{D}<1$, $\E{S\mid D=0}<\E{S\mid D=1}<1$, and $\inf_{u\in\mathbb{S}^{d-1}} V_{\sigma}(u)>0$.
        \item[(iii)] $(Y_1, Y_0, S_1, S_0, D)$ is i.i.d. across individuals.
    \end{description}
\end{assumption}
\begin{proposition}\label{prop:bootstrap}
    Under Assumptions \ref{assumption:setup} and \ref{assumption:bootstrap}, it holds $\lim_{n\to\infty}\mathbb{P}[\mathcal{S}_1 \subset \hat{\mathcal{R}}_1] = 1-\alpha$.
\end{proposition}

In view of the proof of Proposition \ref{prop:bootstrap}, the joint confidence region for $(\Psi(\mu_{0\oplus}), \mathcal{S}_1)$ is also readily available.
Instead of \eqref{eq:stat}, define
\begin{align*}
    T_{\mathtt{joint}}(b) = 
    \max\left\{
    \max_{j\in\{1,\ldots,d\}} \frac{|\tilde{\xi}^{(b)}(j) - \hat{\xi}(j)|}{\sqrt{\hat{V}_{0}(j)/n}},
    \sup_{u\in\mathbb{S}^{d-1}}
    \frac{\tilde{\sigma}_{\mathcal{I}_1}^{(b)}(u) - \hat{\sigma}_{\mathcal{I}_1}(u)}{\sqrt{\hat{V}_{\sigma}(u)/n}}
    \right\},
\end{align*}
where $\hat{\xi}(j)$ and $\tilde{\xi}^{(b)}(j)$ are the sample- and bootstrap- estimators of $\xi(j)\coloneqq$ the $j$-th element of $\E{\Psi(Y)\mid S=1,D=0}$, and ${V}_{0}(j)$ is the asymptotic variance of $\hat{\xi}(j)$ with a consistent estimator $\hat{V}_{0}(j)$.
Then, using the critical value $\hat{\text{j.c.v.}} =$ the $(1-\alpha)$-th quantile of $\{T_{\mathtt{joint}}(b) : b=1,\ldots,B\}$, one can construct a joint confidence region such that
\begin{align}
    \lim_{n\to\infty}\mathbb{P}[\Psi(\mu_{0\oplus}) \in \hat{\mathcal{R}}_{0,\mathtt{joint}}\,\text{ and }\,\mathcal{S}_1 \subset \hat{\mathcal{R}}_{1,\mathtt{joint}}] = 1-\alpha,\label{eq:joint}
\end{align}
where
\begin{align*}
    \hat{\mathcal{R}}_{0,\mathtt{joint}} =
    \left\{
    \zeta\in\mathbb{R}^d : \hat{\xi}(j) - \hat{\text{j.c.v.}} \sqrt{\hat{V}_{0}(j)/n} \leq \zeta(j) \leq  \hat{\xi}(j) + \hat{\text{j.c.v.}} \sqrt{\hat{V}_{0}(j)/n},\,\forall j\in\{1,\ldots,d\}
    \right\}
\end{align*}
and $\hat{\mathcal{R}}_{1,\mathtt{joint}}$ is constructed as before but replacing $\hat{\text{c.v.}}$ with $\hat{\text{j.c.v.}}$.
\begin{corollary}\label{cor:bootstrap}
    In addition to the assumptions made in Proposition \ref{prop:bootstrap}, suppose that $\E{S(1-D)}>0$, $\Psi(Y)\mid (S=1,D=0)$ has a bounded support, and $\min_{j\in\{1,\ldots,d\}} V_0(j)>0$. Then, \eqref{eq:joint} holds.
\end{corollary}
This general construction of the joint confidence region will be useful for conducting a hypothesis test of the treatment effects (see Section \ref{sec:treatment effect}).
When the treatment effect is defined as the difference in the embedded space $\Psi(\mathcal{Y})$, the $1-\alpha$ confidence region for the treatment effect is given by $\hat{\mathcal{R}}_{1,\mathtt{joint}} \ominus \hat{\mathcal{R}}_{0,\mathtt{joint}}$.
If the treatment effect is measured by the projection difference, $\Pi(\Psi^{-1}(\hat{\mathcal{R}}_{1,\mathtt{joint}})) \ominus \Pi(\Psi^{-1}(\hat{\mathcal{R}}_{0,\mathtt{joint}}))$ provides a valid confidence interval. When the geodesic treatment effect is employed, $\{ \gamma_{\nu,\mu} : \nu \in \Psi^{-1}(\hat{\mathcal{R}}_{0,\mathtt{joint}})\,\text{and}\, \mu\in\Psi^{-1}(\hat{\mathcal{R}}_{1,\mathtt{joint}})\}$ has a correct asymptotic coverage.

\section{Empirical illustration}\label{sec:empir}
This section illustrates our proposed method using a distributional outcome: sleep duration (hours slept per night). In particular, we study how individuals' sleep responds to financial incentives. Sleep is a central component of time allocation and a potentially important input into economic performance (e.g., \citealp{Rao_etal:2021, Streatfeild_etal:2021}), and recent work in economics has examined whether incentives can shift sleep behavior (e.g., \citealp{Avery_etal:2025, Bessone_etal:2021}). We use a subset of the data from \cite{Bessone_etal:2021} and reanalyze the experiment with our method, focusing on distributional changes in sleep hours.

In \citet{Bessone_etal:2021}, low-income adults in India were employed in a paid data-entry job and wore an actigraphy-based wearable device throughout the study. After eight baseline days, participants were randomly assigned to one of three arms (control, sleep-aid devices$+$incentives, and sleep-aid devices$+$encouragement) and were further cross-randomized to a workplace nap offer. In the incentives arm, participants received sleep-improving devices and were offered monetary incentives tied to increases in actigraph-measured sleep relative to baseline. In this illustration, we focus on the incentives-versus-control comparison and pool over the nap assignment. As a result, the sample sizes are 152 in the control group and 150 in the treatment (incentives) group.

As the distributional sleep outcome, we use the nightly sleeping hours from days 9-28 and treat the resulting set of hours for each individual as a distributional outcome. For illustration, we classify an individual as a non-respondent if sleep data are missing for at least three nights out of these 20 nights. An assumption here is that, when the number of missing nights is below the threshold, missingness is at random and therefore does not distort the distribution; otherwise, the missingness is potentially endogenous. 
Under this setup, the attrition rate is 12.5\% in the control group and 10.7\% in the treatment group.

\begin{figure}[t]
    \centering
    \includegraphics[width=0.5\linewidth]{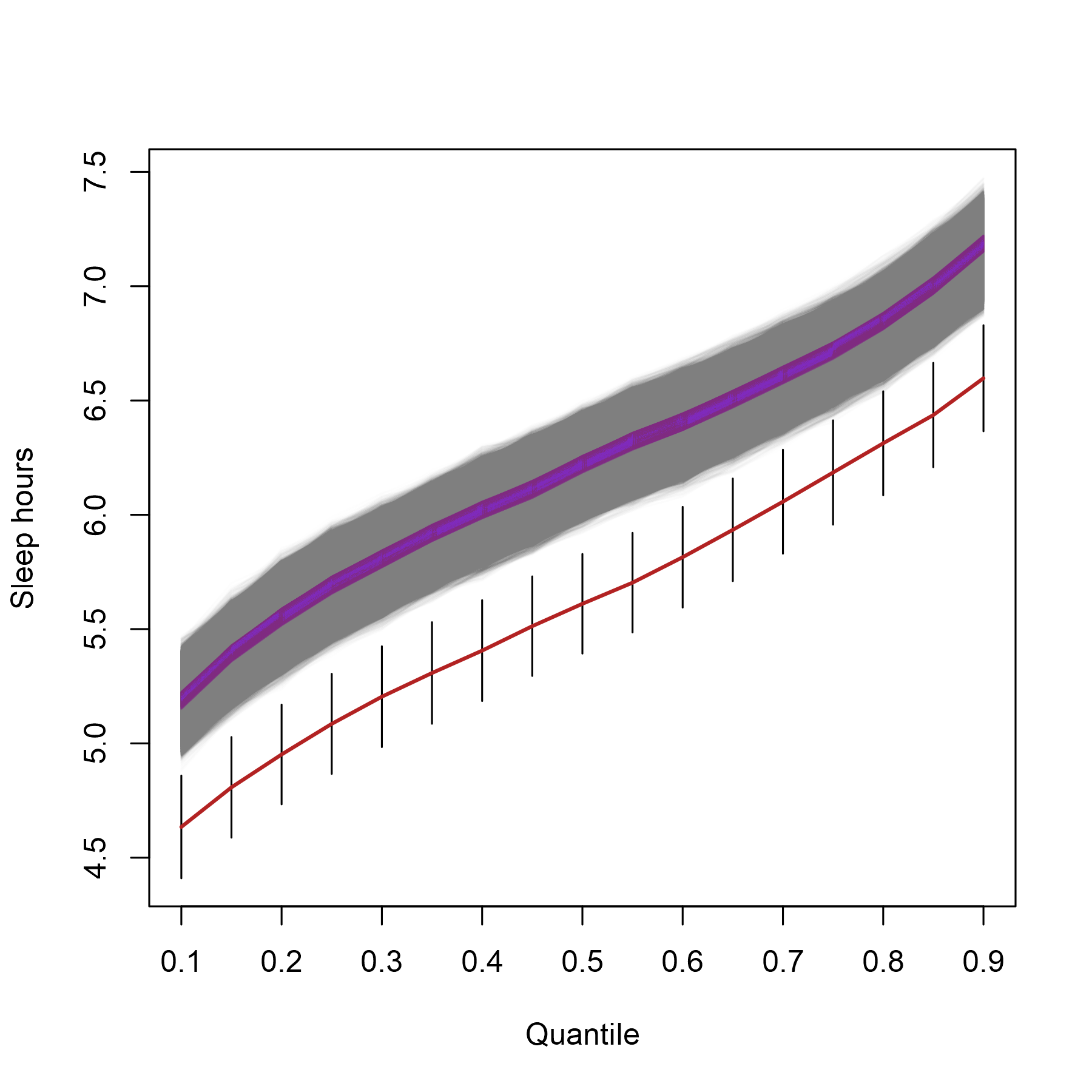}
    \caption{Expected quantile functions}
    \label{fig:quantile}

    \begin{flushleft}
    \footnotesize
    \renewcommand{\baselineskip}{11pt}
    \textbf{Note:} The red line shows the point-identified \f mean for the control group, represented as a quantile function, and the associated vertical bars represent the joint confidence intervals. The inner, purple band is a linearly interpolated approximation of the identified set $\left\{ f \in \Psi(\mathcal{W}) : \bigl(f(q_1),\ldots,f(q_k)\bigr) \in \mathcal{S}_{1,\Pi} \right\}$. The outer, gray region is a linearly interpolated joint confidence region for $\mathcal{S}_{1,\Pi}$.
    \end{flushleft}
\end{figure}

We estimate the identified set following the procedure described in Remark \ref{rem:feasible} and Section \ref{sec:implementation}. 
We choose $k=17$ evaluation points, with $q_1=0.10, q_2=0.15, \ldots, q_{17}=0.90$, and estimate $\mathcal{S}_{1,\Pi}$. We then randomly draw $200{,}000$ points $\{(f(q_1),\ldots,f(q_{17}))_j\}_{j=1}^{200{,}000}$ from $\hat{\mathcal{S}}_{1,\Pi}$, and construct $200{,}000$ linearly interpolated curves connecting $f(q_1),\ldots,f(q_{17})$. The 95\% joint confidence region is computed following Corollary \ref{cor:bootstrap} and visualized in the same manner as $\hat{\mathcal{S}}_{1,\Pi}$.

The estimation results are shown in Figure \ref{fig:quantile}. We observe several notable features. First, the estimated identified set (inner, purple band) is tight and informative. This is because $\hat{p}=(1-0.125)/(1-0.107)\approx 0.98$ is close to one, implying that the observed distribution is only mildly ``contaminated'' and the identified set is therefore close to point identification. Second, the identified set lies above the control mean (red curve), and the joint confidence region shows no overlap between the treatment and control groups. These findings suggest that providing incentives shifts the distribution of sleep upward. Moreover, the observations are consistent with $\mu_{1\oplus}$ first-order stochastically dominating $\mu_{0\oplus}$, which is a richer implication than a comparison of average sleep duration alone.

\newpage

\appendix

\section{Additional examples}\label{app:ex}

\begin{example}[Compositional data allowing zeros]\label{ex:zero}
    If compositional outcomes may contain zero components, the space for compositional data $\mathcal{Y}$ corresponds to the simplex with boundary $\Delta^{k-1}=\{y \in \mathbb{R}^k: y_j \geq 0,j=1,\ldots,k,\text{ and } \, \sum_{j=1}^k y_j = 1\}$.
    In this situation, one can consider the square root transformation $y \mapsto \sqrt{y}=(\sqrt{y_1},\dots,\sqrt{y_k})'$, which allows the compositional outcomes to be regarded as points on the positive orthant of the unit sphere $\mathbb{S}_+^{k-1}=\{x \in \mathbb{R}^k:\|x\|=1, x_j \geq 0, j=1,\dots,k\}$ equipped with the geodesic (Riemannian) metric on the sphere $d_g(x_1,x_2)=\mathrm{arccos}(x'_1x_2)$, $x_1,x_2 \in \mathbb{S}^{k-1}=\{x \in \mathbb{R}^k:\|x\|=1\}$ \citep{scea:11,scea:14}. This approach incorporates the geometric structure of compositional data allowing zeros in a manner analogous to the Aitchison metric. Choose a reference point $\mu \in \mathbb{S}_+^{k-1}$ and let $T_\mu\mathbb{S}^{k-1}=\{x \in \mathbb{R}^k: \mu'x=0\}$ be the tangent space at $\mu$. A typical choice of $\mu$ is $(1/\sqrt{3},1/\sqrt{3},1/\sqrt{3})$. Using the logarithmic map $\mathrm{Log}_\mu: \mathbb{S}^{k-1}\backslash\{-\mu\} \to T_\mu \mathbb{S}^{k-1}$, $\mathrm{Log}_\mu(x)=\mathrm{arccos}(x'\mu)\frac{(I_k-\mu\mu')(x-\mu)}{\|(I_k-\mu\mu')(x-\mu)\|}$, the compositional outcomes can be mapped into a closed convex subset $\mathrm{Log}_\mu(\mathbb{S}_+^{k-1})$ of a vector space $T_\mu\mathbb{S}^{k-1}$. Note that the inverse map of $\mathrm{Log}_\mu$ is given by the exponential map $\mathrm{Exp}_\mu: T_\mu\mathbb{S}^{k-1} \to \mathbb{S}^{k-1}$, $\mathrm{Exp}_\mu(v)=\cos(\|v\|)\mu + \sin(\|v\|)\frac{v}{\|v\|}$ so that $\mathrm{Exp}_\mu(\mathrm{Log}_\mu(\mathbb{S}_+^{k-1}))=\mathbb{S}_+^{k-1}$.
\end{example}

\begin{example}[Multivariate probability distributions]\label{ex:multi-dist}
  Unlike the case where the outcome is a univariate probability distribution, when the outcome is given as a multivariate probability distribution, it is not straightforward to define a transformation analogous to the quantile function. However, by introducing an alternative transformation—motivated by the Aitchison metric for compositional data—for multivariate probability density functions, it becomes possible to embed a collection of multivariate probability distributions isometrically into a separable Hilbert space. Let $(\Omega, \mathcal{A})$ be a measurable space with $\Omega \subset \mathbb{R}^d$ and $\mu$ be a probability measure on $\Omega$. Define $M(\mu)$ as the set of measures on $(\Omega, \mathcal{A})$ that are $\sigma$-finite and mutually absolutely continuous with respect to $\mu$. From the definition of $M(\mu)$, any measure $\eta \in M(\mu)$ admits a Radon-Nikodym derivative (density function) with respect to $\mu$ denoted by $\frac{d\eta}{d\mu}$ such that $\eta(A) = \int_A \frac{d\eta}{d\mu}(x)d\mu(x)$. Note that the function $\frac{d\eta}{d\mu}$ is unique $\mu$ almost everywhere. Define an equivalence relation $=_B$ between two measures $\eta, \nu \in M(\mu)$ as $\eta =_B \nu$ if and only if there exists a constant $c>0$ such that $\frac{d\eta}{d\mu} = c\frac{d\nu}{d\mu}$ $\mu$-almost everywhere. Define $B(\mu)$ as the set of equivalence classes with respect to the relation $=_B$ within $M(\mu)$. Consider the space
    \[
    B^2(\mu) = \left\{\eta \in B(\mu): \int_{\Omega} \left(\log \left[{d\eta \over d\mu}(x)\right]\right)^2d\mu(x)<\infty\right\}
    \]
    equipped with distance $d_{\mathrm{BH}}(\eta,\nu) = \sqrt{\int_\Omega \left\{[\Psi(\eta)](x)-[\Psi(\nu)](x)\right\}^2d\mu(x)}$, where $[\Psi(\eta)](x)=\log \left[\frac{d\eta}{d\mu}(x)\right] - \int_{\Omega} \log \left[\frac{d\eta}{d\mu}(x)\right]d\mu(x)$. Note that $\eta \in B^2(\mu)$ if and only if $\log \left[\frac{d\eta}{d\mu}\right] \in L^2(\mu)$ where $L^2(\mu)$ is the space of square integrable functions on $\Omega$ with respect to $\mu$. The space of multivariate densities $(B^2(\mu),d_{\mathrm{BH}})$ is called the Bayes Hilbert space \citep{Boogaart_etal:2014} and the image space $\Psi(B^2(\mu))=\{g \in L^2(\mu): \int_\Omega g(x)d\mu(x)=0\}$ is a closed convex subset of the separable Hilbert space $L^2(\mu)$. 
\end{example}

\begin{example}[Symmetric positive semidefinite matrices]
    Let $\text{Sym}_m^{+}$ denote the space of $m \times m$ symmetric positive semidefinite matrices, and $\text{Sym}_m^{++}$ the space of $m \times m$ symmetric positive definite matrices.
    These spaces have been studied for applications to neuroimaging \citep{dryden2009non}, signal processing \citep{arna:13}, and linguistic analysis \citep{pigoli2014distances}. A natural starting point is the Frobenius metric $d_F$ on $\text{Sym}_m^+$. In this case, the space $\text{Sym}_m^+$ forms a closed convex subset of the Euclidean space $\mathbb{R}^{m^2}$. For any constant $p > 0$, the power metric $d_{F, p}$ \citep{dryden2009non} on $\text{Sym}_m^+$ is defined as  $d_{F, p}(A, B) = d_F(A^p, B^p)$ where $A^p = U\Lambda^p U'$ and $U\Lambda U'$ is the usual spectral decomposition of the matrix $A$. In this case, the mapping $\Psi(A) = A^p$ is an isometry from $\text{Sym}_m^+$ to the Euclidean space $\mathbb{R}^{m^2}$, and the image $\Psi(\text{Sym}_m^+) = \text{Sym}_m^+$ is closed and convex in $\mathbb{R}^{m^2}$.
    The Log-Euclidean metric $d_{\mathrm{LE}}$ \citep{arsigny2007geometric} on $\text{Sym}_m^{++}$ is defined as $d_{\mathrm{LE}}(A,B) = d_F(\log(A),\log(B))$, where $\log(A)$ denotes the matrix logarithm of the matrix $A$.
    In this case, the mapping $\Psi(A) =\log(A)$ is an isometry from $\text{Sym}_m^{++}$ to the Euclidean space $\mathbb{R}^{m^2}$, and the image $\Psi(\text{Sym}_m^{++}) = \{A \in \mathbb{R}^{m^2}: A = A'\}$ is closed and convex in $\mathbb{R}^{m^2}$. 
\end{example}

\section{Proofs}\label{app:pf}
\subsection{Sharpness of $\mathcal{I}_1$}\label{sec:sharp}
We begin by defining the sharpness of an identified set. The argument that follows is borrowed from \cite{Canay_Shaikh:2017}.
Suppose that a researcher observes data with distribution $P\in\mathcal{P}\coloneqq\{P_\gamma:\gamma\in\Gamma\}$, where the set $\mathcal{P}$ is characterized by a possibly infinite-dimensional parameter $\gamma\in\Gamma$. Define $\Gamma_0(P)\coloneqq \{\gamma\in\Gamma : P_\gamma = P\}$.
Suppose we are interested in a parameter $\theta = \theta(\gamma)$. The sharp identified set of $\theta$ is defined as follows.
\begin{definition}\label{def}
    The sharp identified set of $\theta$ is defined as $\theta(\Gamma_0(P)) = \{\theta(\gamma) : \gamma \in \Gamma_0(P)\}$.
\end{definition}
That is, $\theta(\Gamma_0(P))$ is sharp in the sense that for any value $\theta_1 \in \theta(\Gamma_0(P))$, there exists $\gamma_1$ such that $\theta_1 = \theta(\gamma_1)$ and $P_{\gamma_1} = P$, i.e., consistent with observed data distribution.
Any superset of $\theta(\Gamma_0(P))$ is called a (non-sharp) valid identified set.

Recall the setting introduced in Section \ref{sec:gen}.
Let $\mathcal{P}(\mathcal{H})$ deonote the set of probability distributions on $\mathcal{H}$ and $\mathcal{M}_+(\mathcal{H})$ denote the set of nonnegative measures on $\mathcal{H}$. 
In this setup, we can define
\begin{align*}
&\Gamma = \{\gamma=(\nu,\eta): \nu, \eta \in  \mathcal{M}_+(\mathcal{H}),\ \nu+\eta \in \mathcal{P}(\mathcal{H}),\ \nu(\mathcal{H})=p\},\\
&\mathcal{P}=\{P_\gamma: \gamma \in \Gamma\},\ P_\gamma= \nu + \eta,\ P =\mu,\\
&\Gamma_0(P) = \{\gamma \in \Gamma: P_{\gamma} = P\},\quad \theta(\gamma) = {1 \over p}\int_\mathcal{H}x\,d\nu(x). 
\end{align*}
Then the sharp identified set is $\{\theta(\gamma): \gamma \in \Gamma_0(P)\}$, which can be rewritten as
\begin{align*}
    &\{\theta(\gamma): \gamma \in \Gamma_0(P)\} \\
    &= 
    \left\{{1 \over p}\int_\mathcal{H}x\,d\nu(x): \nu,\eta\in\mathcal{M}_+(\mathcal{H}),\, \nu(\mathcal{H})=p, \,\nu+\eta=\mu\right\}\\
    &= 
    \left\{{1 \over p}\int_\mathcal{H}x\,d\nu(x): \nu\in\mathcal{M}_+(\mathcal{H}),\, \nu(\mathcal{H})=p,\, \nu(\mathcal{A})\leq\mu(\mathcal{A})\text{ for any measurable } \mathcal{A}\subset\mathcal{H} \right\} = \mathcal{I}_1.
\end{align*}
Hence, $\mathcal{I}_1$ is the sharp identified set, as summarized below.
\begin{lemma}\label{lemma:sharp}
    Suppose Assumptions \ref{assumption:setup} and \ref{assumption:metric} hold true. Then, $\mathcal{I}_1$ in \eqref{eq:sharpset} is the sharp identified set of $\E{\Psi(Y_1) \mid S_1=S_0=1}$.
\end{lemma}

\subsection{Proof of Theorem \ref{thm}}
\subsubsection*{Proof of (i)}
We shall characterize the set $\overline{\mathrm{conv}}\, \mathcal{I}_1$ by using the support function. For $\mathcal{A} \subset \mathcal{H}$, let $\sigma_{\mathcal{A}}(\cdot)=\sup_{v\in \mathcal{A}}\langle \cdot,v \rangle$ be its support function defined on the unit sphere of $\mathcal{H}$. By the property of the support function, the convex closure of $\mathcal{I}_1$ admits the following representation (\citealp[Proposition 7.11]{Bauschke_Combettes:2017}):
\[
    \overline{\mathrm{conv}}\,\mathcal{I}_1 = \bigcap_{\substack{u\in\mathcal{H}}}\left\{v\in\mathcal{H} : \langle u, v \rangle \leq \sigma_{\mathcal{I}_1}(u)\right\}.
\]
Noting that $\{v\in\mathcal{H} : \langle cu, v \rangle \leq \sigma_{\mathcal{I}_1}(cu)\} = \{v\in\mathcal{H} : \langle u, v \rangle \leq \sigma_{\mathcal{I}_1}(u)\}$ for any $c\geq0$, it suffices to consider $u$ such that $\|u\|_{\mathcal{H}}=1$, that is,
\[
    \overline{\mathrm{conv}}\,\mathcal{I}_1 = \bigcap_{\substack{u\in\mathcal{H}\\ \|u\|=1}}\left\{v\in\mathcal{H} : \langle u, v \rangle \leq \sigma_{\mathcal{I}_1}(u)\right\}.
\]
We will derive an alternative representation of the support function of $\mathcal{I}_1$.
Recall that, for any $v \in \mathcal{I}_1$, there exists a finite measure $\nu$ such that $0\leq \nu\leq\mu$ (thereby $\nu\ll\mu$), $\nu(\mathcal{H})=p$, and $v = (1/p) \int x\,d\nu(x)$.
Then, with $f=d\nu/d\mu\in[0,1]$, we have that
\begin{align*}
    \langle u,v \rangle = \frac{1}{p}\int_\mathcal{H} \langle u,x \rangle\,d\nu(x)
    = \frac{1}{p}\int_\mathcal{H} \langle u,x \rangle f(x)\,d\mu(x),
\end{align*}
so that we have that
\begin{align*}
    \sigma_{\mathcal{I}_1}(u) &= 
    \sup_{v\in\mathcal{I}_1} \langle u,v \rangle
    =\sup_{0\leq f \leq 1, \int f\,d\mu = p}\frac{1}{p}\int_\mathcal{H} \langle u,x \rangle f(x)\,d\mu(x)\\
    &=\frac{1}{p} \int_\mathcal{H} \langle u,x \rangle \mathbf{1}\{\langle u,x \rangle \geq Q_{\langle u, \Psi(Y)\rangle}(1-p)\}\,d\mu(x)\\
    &=\E{\langle u, \Psi(Y)\rangle\mid  D = 1, S = 1, \langle u, \Psi(Y)\rangle \geq Q_{\langle u, \Psi(Y)\rangle}(1-p)},
\end{align*}
where $Q_{\langle u, \Psi(Y)\rangle}(1-p)$ denotes the $(1-p)$-th conditional quantile of $\langle u, \Psi(Y)\rangle$ given $D = 1, S = 1$.
The third equality uses the continuity of the conditional distribution of $\langle u, \Psi(Y)\rangle$.
This shows the first assertion. The validity of $\mathcal{S}_1 = \overline{\mathrm{conv}}\,  \mathcal{I}_1$ follows by observing that $\mathcal{I}_1\subseteq \overline{\mathrm{conv}}\, \mathcal{I}_1$.

\subsubsection*{Proof of (ii)}
For every $y\in \Psi^{-1}(\mathcal{S}_1)$, $\Psi(y)\in\mathcal{S}_1$ holds, implying the validity of the bounds.

\subsubsection*{Proof of (iii)}
It suffices to show that $\mathcal{I}_1$ is closed convex when $\Psi(Y)$ is finite dimensional (or $\Psi(Y)\in\mathbb{R}^k$).
To see this, we introduce $\mathcal{B}_1 \coloneqq \left\{\E{\Psi(Y)\mid D = 1, S = 1, \Psi(Y)\in \mathcal{A}}:\mathcal{A}\in\mathscr{A}\right\}$, where $\mathscr{A}$ is the collection of all subsets $\mathcal{A}$ such that $\P{\Psi(Y)\in \mathcal{A} \mid  D = 1, S = 1} = p$.
We prove that $\mathcal{I}_1 = \mathcal{B}_1 = \overline{\mathrm{conv}}\, \mathcal{B}_1$ when $\Psi(Y)$ is finite dimensional.

We begin by showing that $\mathcal{B}_1 = \overline{\mathrm{conv}}\, \mathcal{B}_1$.
Define the $\mathbb{R}^{k+1}$-valued vector measure $m$ by
\begin{align*}
    m(\mathcal{A}) \coloneqq \left(\mu(\mathcal{A}), \int_\mathcal{A} x\,d\mu(x)\right),
\end{align*}
where $\mathcal{A}$ is a Borel subset. Since $m$ is a finite, non-atomic measure, the Lyapunov convexity theorem (e.g., \citealp[Theorem 13.33]{Aliprantis_Border:2006}) shows that the range
\begin{align*}
    \mathcal{R}(m)\coloneqq \left\{m(\mathcal{A}): \mathcal{A} \in \text{the Borel subset of } \mathbb{R}^k\right\}
\end{align*}
is closed and convex, so is the slice $\mathcal{R}_p \coloneqq \{y\in\mathbb{R}^k : (p,y) \in \mathcal{R}(m)\}$.
Then, we have that $\mathcal{B}_1 = \{(1/p) \int_\mathcal{A} x\,d\mu(x) : \mathcal{A}\in\mathscr{A}\} = \mathcal{R}_p / p$ is closed and convex, i.e., $\mathcal{B}_1 = \overline{\mathrm{conv}}\,\mathcal{B}_1$.

Next, we see that $\mathcal{I}_1=\mathcal{B}_1$.
$\mathcal{I}_1\supset\mathcal{B}_1$ is immediate, so we show the reverse inclusion, $\mathcal{I}_1\subset\mathcal{B}_1$.
Take any $v\in\mathcal{I}_1$, then there is a finite measure $\nu$ with Radon-Nikodym derivative $f$ as in the proof of (i).
Note that
\begin{align*}
    \left(p, \int_\mathcal{\mathcal{H}} x\,d\nu(x)\right)&= 
    \left(\int_\mathcal{\mathcal{H}} f(x)\,d\mu(x), \int_\mathcal{\mathcal{H}} x f(x)\,d\mu(x)\right) \\
    &=
    \int_0^1\left(\mu(\{f\geq t\}) , \int_{\{f\geq t\}} x\,d\mu(x)\right)\,dt = \int_0^1 m(\{f\geq t\})\,dt,
\end{align*}
where the second equality uses the representation $f(x) = \int_0^1 \mathbf{1}\{f(x)\geq t\}\,dt$ and Fubini's theorem.
Since $\mathcal{R}(m)$ is closed and comvex, $\int_0^1 m(\{f\geq t\})\,dt \in \mathcal{R}(m)$ so that there exists a measurable set $\mathcal{A}_0$ such that $m(\mathcal{A}_0) = \left(p, \int_\mathcal{\mathcal{H}} x\,d\nu(x)\right)$, or $\mu(\mathcal{A}_0)=p$ and $\int_{\mathcal{A}_0} x\,d\mu(x) = \int_\mathcal{\mathcal{H}} x\,d\nu(x)$. Consequently,
\begin{align*}
    v = \frac{1}{p}\int_{\mathcal{H}} x\,d\nu(x) = \frac{1}{\mu(\mathcal{A}_0)}\int_{\mathcal{A}_0} x\,d\mu(x)\in \mathcal{B}_1.
\end{align*}
Therefore, we have $\mathcal{I}_1\subset\mathcal{B}_1$ and hence $\mathcal{I}_1=\mathcal{B}_1$.

To sum up, we have shown that $\mathcal{I}_1$ is closed and convex.
Then we obtain that $\mathcal{I}_1 = \overline{\mathrm{conv}}\,\mathcal{I}_1 = \mathcal{S}_1$, i.e., $\mathcal{S}_1$ is the sharp identified set.
The sharpness of $\Psi^{-1}(\mathcal{S}_1)$ holds as follows.
If $\exists y\in \Psi^{-1}(\mathcal{S}_1)$ such that this $y$ is not attainable as $\mu_{1\oplus}$, this implies that $\mathcal{S}_1$ contains a non-attainable point $\Psi(y)$, contradicting the sharpness of $\mathcal{S}_1$. Hence, $\Psi^{-1}(\mathcal{S}_1)$ is also sharp.\hfill$\clubsuit$

\subsection{Proof of Proposition \ref{prop:bootstrap}}
Below, we write $a \lesssim b$ to mean that there exists a constant $C\in(0,\infty)$ such that $a \leq Cb$.
\subsubsection*{Step 1}
In this part, we rewrite our estimator as a $Z$-estimator and outline the bootstrap validity, while some technical argument is postponed to \textit{Step 2}.
To simplify the notation, let $Z_u \coloneqq \langle u, \Psi(Y)\rangle$, and define $F_u$ as the conditional distribution function of $Z_u$ given $S=1, D=1$ with a conditional density function $f_u$.

We can write that $\sigma_{\mathcal{I}_1}(u)= \E{Z_u \mid  D = 1, S = 1, Z_u \geq Q_{Z_u}(1-p)}$.
Let $\theta_u\coloneqq (\theta_1(u),\theta_2(u),\theta_3, \theta_4)^\top$ and $\theta_u^*\coloneqq (\theta_1^*(u),\theta_2^*(u),\theta_3^*,\theta_4^*)^\top$, where
\begin{align*}
    \theta_1^*(u) \coloneqq \sigma_{\mathcal{I}_1}(u),\quad
    \theta_2^*(u) \coloneqq Q_{Z_u}(1-p),\quad
    \theta_3^* \coloneqq p,\quad \text{and }
    \theta_4^* \coloneqq \P{S=1 \mid D=0}.
\end{align*}
Here, we define $\Omega(\theta_u)\coloneqq \E{\psi_{\theta,u}} (=\mathbb{P} \psi_{\theta,u} \text{ in the empirical process notation}) \in \ell^\infty(\mathbb{S}^{d-1})^2 \times \mathbb{R}^2$, where
\begin{align*}
    \psi_{\theta,u}(Y,S,D) \coloneqq 
    \begin{pmatrix}
         {\left(Z_u - \theta_1(u)\right) SD \mathbf{1}\left\{Z_u \geq \theta_2(u)\right\}}\\
        {\left(\mathbf{1}\left\{Z_u \geq \theta_2(u)\right\} - \theta_3\right)S D}\\
        {\left(\theta_3 S - \theta_4\right)D}\\
        {(S-\theta_4)(1-D)}
    \end{pmatrix}.
\end{align*}
Let $\Omega_n (= \mathbb{P}_n \psi_{\theta,u})$ be the sample analog of $\Omega$, and the estimator $(\hat{\sigma}_{\mathcal{I}_1}(u), \hat{Q}_{Z_u}(1-\hat{p}), \hat{p}, \hat{\mathbb{P}}[S=1\mid D=0])^\top$ is written by $\hat{\theta}_u\coloneqq (\hat{\theta}_1(u),\hat{\theta}_2(u),\hat{\theta}_3,\hat{\theta}_4)^\top$.
Then, we can see that $\Omega(\theta_u^*)=0$ and $\Omega_n(\hat{\theta}_u)=0$.
The bootstrap counterparts are denoted by $\tilde{\theta}_u$ and $\tilde{\Omega}_n$, which satisfy that $\tilde{\Omega}_n(\tilde{\theta}_u)=0$.

The representation above suggests that $\hat{\theta}_u$ (and more specifically  $\hat{\sigma}_{\mathcal{I}_1}(u)$) is a $Z$-estimator.
Using this structure, we begin with an asymptotic approximation of the non-studentized statistic.

To this aim, define
\begin{align*}
    \mathcal{F}\coloneqq \left\{(Y,S,D)\mapsto \psi_{\theta,u}(Y,S,D) : 
    \theta \in \Theta, u \in \mathbb{S}^{d-1}
    \right\},
\end{align*}
where $\Theta$ denotes a parameter space:
\begin{align*}
    \Theta = \left\{\theta\in\ell^\infty(\mathbb{S}^{d-1})^2 \times \mathbb{R}^2 : \sup_{u\in \mathbb{S}^{d-1}} |\theta_1(u)| < M, \sup_{u\in \mathbb{S}^{d-1}} |\theta_2(u)| < M, \theta_3\in[0,1],\text{ and }\theta_4\in[0,1]\right\},
\end{align*}
equipped with the norm $\|\theta\| \coloneqq \sup_{u\in \mathbb{S}^{d-1}} |\theta_1(u)| + \sup_{u\in \mathbb{S}^{d-1}} |\theta_2(u)| + |\theta_3| + |\theta_4|$.
Note that one can take such an $M$ by Assumption \ref{assumption:bootstrap} (i).
We shall verify conditions (A) to (F) in Theorem 13.4 of \citet[p.~256]{Kosorok:2008} so that we obtain $\sqrt{n}(\hat{\theta}_u - \theta_u^*) \rightsquigarrow -\dot{\Omega}_{\theta_u^*}^{-1} Z$ and $\sqrt{n}(\tilde{\theta}_u - \hat{\theta}_u) \underset{W}{\overset{\mathbb{P}}{\rightsquigarrow}} -\dot{\Omega}_{\theta_u^*}^{-1} Z$, where $W$ indicates the multinomial weights given the data, and $Z$ is the tight mean zero Gaussian limiting distribution of $\sqrt{n}(\Omega_n - \Omega)(\theta^*)$.
Condition (E) is satisfied by construction. We will show the remaining conditions below.

We begin with condition (A).
Let $\{\theta_n\}\subset\Theta$ satisfy $\|\Omega(\theta_n)\|\to 0$.
From the fourth component, $\mathbb{E}[(S-\theta_{4,n})(1-D)]=o(1)$ yields $\theta_{4,n}\to \theta_4^*=\mathbb{P}(S=1\mid D=0)$.
From the third component, $\mathbb{E}[(\theta_{3,n}S - \theta_{4,n})D]=o(1)$ gives $\theta_{3,n}\to \theta_3^*$ since $\mathbb{E}[SD]>0$ and $\theta_{4,n}\to \theta_4^*$.
Next, the second component implies $\sup_{u\in\mathbb{S}^{d-1}}\left|\E{\left(\mathbf{1}\{Z_u\ge \theta_{2,n}(u)\}-\theta_{3,n}\right)SD}\right| = o(1)$, and hence $\mathbb{P}(Z_u\ge \theta_{2,n}(u)\mid S=1,D=1)=\theta_{3,n}+o(1)$, or we have $\sup_{u\in\mathbb{S}^{d-1}}\left|F_u(\theta_{2,n}(u)) - F_u(\theta_2^*(u))\right|\to0$.
The bound derived in \textit{Step 2} implies that $\inf_{u\in\mathbb{S}^{d-1}} \inf_{t:|t-\theta_2^*(u)|<\eta} f_u(t) \geq \underline{f}_Z>0$ for some $\eta>0$. 
We have $\sup_{u\in\mathbb{S}^{d-1}} |\theta_{2,n}(u) - \theta_2^*(u)| < \eta$ for $n$ large enough (otherwise $\left|F_u(\theta_{2,n}(u)) - F_u(\theta_2^*(u))\right|\geq \underline{f}_Z \eta$ for some $u$, a contradiction).
Hence, the mean value theorem suggests that 
\begin{align*}
    |\theta_{2,n}(u)-\theta_2^*(u)| \leq 
    \frac{1}{\underline{f}_Z} \left|F_u(\theta_{2,n}(u)) - F_u(\theta_2^*(u))\right|,
\end{align*}
which yields
\begin{align*}
    \sup_{u\in\mathbb{S}^{d-1}}|\theta_{2,n}(u)-\theta_2^*(u)| \leq 
    \frac{1}{\underline{f}_Z} \sup_{u\in\mathbb{S}^{d-1}}\left|F_u(\theta_{2,n}(u)) - F_u(\theta_2^*(u))\right|\to0.
\end{align*}
Similarly, we obtain that $\sup_{u\in\mathbb{S}^{d-1}}|\theta_{1,n}(u)-\theta_1^*(u)|\to 0$.
Therefore $\|\theta_n-\theta^*\|\to 0$, establishing condition (A).

We next treat conditions (B) and (C).
Note that we can take $\mathcal{E}$ such that
$\mathcal{F} \subset \mathcal{E}$ and 
\begin{align*}
    \mathcal{E} \coloneqq \left\{\phi_{\alpha, u} : 
    \alpha\coloneqq (\alpha_1,\ldots,\alpha_4)^\top \in V, u \in \mathbb{S}^{d-1}
    \right\},
\end{align*}
where $V\subset\mathbb{R}^4$ is some compact set, and
\begin{align*} 
    \phi_{\alpha, u}(Y,S,D) \coloneqq
    \begin{pmatrix}
        \phi_{\alpha, u}^{(1)}(Y,S,D)\\
        \phi_{\alpha, u}^{(2)}(Y,S,D)\\
        \phi_{\alpha, u}^{(3)}(Y,S,D)\\
        \phi_{\alpha, u}^{(4)}(Y,S,D)
    \end{pmatrix} \coloneqq
    \begin{pmatrix}
        {\left(Z_u - \alpha_1\right) SD \mathbf{1}\left\{Z_u \geq \alpha_2\right\}}\\
        {\left(\mathbf{1}\left\{Z_u \geq \alpha_2\right\} - \alpha_3\right)S D}\\
        {\left(\alpha_3 S - \alpha_4\right)D}\\
        {(S-\alpha_4)(1-D)}
    \end{pmatrix}.
\end{align*}
By the forms of $\phi_{\alpha, u}^{(j)}$, the class $\mathcal{E}$ is a VC-subgraph class. 
Indeed, the indicator class $\{(Y,S,D) \mapsto \mathbf{1}\{\langle u, \Psi(Y) \rangle \geq t\} : u\in\mathbb{S}^{d-1}, t\in\mathbb{R}\}$ corresponds to halfspaces in $\mathbb{R}^d$ and is therefore VC class. Multiplication by the bounded binary variables and by functions from a finite-dimensional linear class preserves the VC property under the standard closure result, and hence $\mathcal{E}$ is VC.
By Assumption \ref{assumption:bootstrap} (i), its envelope has a finite second moment. Hence $\mathcal{E}$ is Donsker, and therefore Glivenko-Cantelli, implying that so is $\mathcal{F}$.
This shows that condition (B) holds. Condition (C) also readily follows.

Next, we show that condition (D) is met. 
In particular, we show that
\begin{align*}
    \sup_{u\in\mathbb{S}^{d-1}}\sum_{j=1}^{4} \E{\left|\Delta_{\theta, u}^{(j)}\right|^2}\to 0\quad \text{ as } \, \theta\to\theta^*,
\end{align*}
where $\Delta_{\theta,u}^{(j)} \coloneqq \psi_{\theta, u}^{(j)} - \psi_{\theta^*, u}^{(j)}$.
It is straightforward to see $\mathbb{E}[|\Delta_{\theta,u}^{(4)}|^2]\to 0$ and $\mathbb{E}[|\Delta_{\theta,u}^{(3)}|^2]\to 0$.
For $j=2$, note that
\begin{align*}
    \left|\Delta_{\theta,u}^{(2)}\right|^2 &\leq 
    \left\{\left|\mathbf{1}\left\{Z_u \geq \theta_2(u)\right\} - \mathbf{1}\left\{Z_u \geq \theta_2^*(u)\right\}\right|SD + \left|\theta_3^* - \theta_3\right|\right\}^2\\
    &\leq 
    2\left|\mathbf{1}\left\{Z_u \geq \theta_2(u)\right\} - \mathbf{1}\left\{Z_u \geq \theta_2^*(u)\right\}\right|^2SD + 2\left|\theta_3^* - \theta_3\right|^2.
\end{align*}
Then, it suffices to show that 
\begin{align*}
    \sup_{u\in\mathbb{S}^{d-1}}\P{\min\{\theta_2(u), \theta_2^*(u)\} \leq Z_u \leq \max\{\theta_2(u), \theta_2^*(u)\}\mid S=1, D=1}\to0.
\end{align*}
It indeed holds by observing that
\begin{align*}
    &\sup_{u\in\mathbb{S}^{d-1}}\P{\min\{\theta_2(u), \theta_2^*(u)\} \leq Z_u \leq \max\{\theta_2(u), \theta_2^*(u)\}\mid S=1, D=1} \\
    &= 
    \sup_{u\in\mathbb{S}^{d-1}}\left|\int_{\min\{\theta_2(u), \theta_2^*(u)\}}^{\max\{\theta_2(u), \theta_2^*(u)\}} f_u(x)\,dx\right| 
    \leq \sup_{u\in\mathbb{S}^{d-1}, x\in \mathbb{R}} f_{u}(x) \cdot \sup_{u\in\mathbb{S}^{d-1}}|\theta_2(u) - \theta_2^*(u)|,
\end{align*}
where the first part in the right-hand side is bounded by \textit{Step 2} and the second part converges to zero by assumption.
The case $j=1$ can be treated similarly:
Note that
\begin{align*}
    \Delta_{\theta,u}^{(1)}
    =\psi_{\theta,u}^{(1)}-\psi_{\theta^*,u}^{(1)}
    =\left(Z_u-\theta_1(u)\right)SD\mathbf 1\{Z_u\ge \theta_2(u)\}
    -\left(Z_u-\theta_1^*(u)\right)SD\mathbf 1\{Z_u\ge \theta_2^*(u)\}.
\end{align*}
By adding and subtracting $(Z_u-\theta_1^*(u))SD\mathbf 1\{Z_u\ge \theta_2(u)\}$, we obtain
\begin{align*}
    \Delta_{\theta,u}^{(1)}
    &=SD\left(\theta_1^*(u)-\theta_1(u)\right)\mathbf 1\{Z_u\ge \theta_2(u)\}
    +SD\left(Z_u-\theta_1^*(u)\right)\left(\mathbf 1\{Z_u\ge \theta_2(u)\}-\mathbf 1\{Z_u\ge \theta_2^*(u)\}\right).
\end{align*}
Then, we have
\begin{align*}
\mathbb E\left[\left|\Delta_{\theta,u}^{(1)}\right|^2\right]
&\le 2\,|\theta_1(u)-\theta_1^*(u)|^2\,\mathbb E\!\left[SD\,\mathbf 1\{Z_u\ge \theta_2(u)\}\right]  \\
&\quad +2\,\mathbb E\!\left[SD\,(Z_u-\theta_1^*(u))^2
\left|\mathbf 1\{Z_u\ge \theta_2(u)\}-\mathbf 1\{Z_u\ge \theta_2^*(u)\}\right|^2\right] \\
&\le 2\,|\theta_1(u)-\theta_1^*(u)|^2\\
&\quad+2\,\mathbb E\!\left[SD\,(Z_u-\theta_1^*(u))^2
\mathbf 1\left\{\min\{\theta_2(u), \theta_2^*(u)\}\leq Z_u\leq \max\{\theta_2(u),\theta_2^*(u)\}\right\}\right].
\end{align*}
The first term converges to $0$.
By Assumption \ref{assumption:bootstrap} (i), $\sup_{u\in\mathbb S^{d-1}} |Z_u| \lesssim 1$ and $\sup_{u\in\mathbb S^{d-1}}|Z_u-\theta_1^*(u)|^2\lesssim 1$ almost surely on $\{S=1,D=1\}$.
Hence, the second term converges to $0$ uniformly in $u$ by the argument used in the case $j=2$, which verifies condition (D) for the first component.

Finally, we verify condition (F).
Let $\Omega^{(j)},j\in\{1,\ldots,4\}$ denotes the $j$-th component of $\Omega$.
Define the linear operator $\dot\Omega_{\theta^*}:\Theta\to
\ell^\infty(\mathbb S^{d-1})^2\times\mathbb R^2$ by
\begin{align*}
    \dot\Omega_{\theta^*}^{(1)}(h)(u)
    &\coloneqq -\E{SD\,\mathbf 1\{Z_u\ge \theta_2^*(u)\}}\,h_1(u)
    -\E{SD}\left(\theta_2^*(u)-\theta_1^*(u)\right)f_u(\theta_2^*(u))\,h_2(u),\\
    \dot\Omega_{\theta^*}^{(2)}(h)(u)
    &\coloneqq -\E{SD}\,f_u(\theta_2^*(u))\,h_2(u)
    -\E{SD}\,h_3,\\
    \dot\Omega_{\theta^*}^{(3)}(h)
    &\coloneqq \E{SD}\,h_3- \E{D}\,h_4,\,\,\text{ and }\,\,
    \dot\Omega_{\theta^*}^{(4)}(h)
    \coloneqq -\E{1-D}\,h_4.
\end{align*}
By a similar argument used in the verification of conditions (A) and (D),
there exists a function $\omega(\delta)\downarrow 0$ such that, whenever $\|\theta-\theta^*\|\leq\delta$,
\begin{align*}
    \sup_{u\in\mathbb S^{d-1}}
    \big|\Omega^{(j)}(\theta)(u)-\Omega^{(j)}(\theta^*)(u)
      -\dot\Omega_{\theta^*}^{(j)}(\theta-\theta^*)(u)\big|
    \lesssim \omega(\delta)\,\|\theta-\theta^*\|,\quad j=1,2,
\end{align*}
and similarly for $j=3,4$, where the dependence on$u$ disappears.
This implies that $\Omega$ is \f differentiable at $\theta^*$ with derivative
$\dot\Omega_{\theta^*}$.

It remains to show that $\dot\Omega_{\theta^*}$ is continuously invertible, i.e., that its inverse is a bounded linear operator.
Assumption \ref{assumption:bootstrap} (ii) and \textit{Step 2} implies that all denominators appearing in the explicit solution to $\dot\Omega_{\theta^*}(h) = g$, where $g\in\ell^\infty(\mathbb S^{d-1})^2\times\mathbb R^2$, are uniformly bounded away from zero over $u\in\mathbb S^{d-1}$, which implies that $\|h\| \lesssim \|g\|$ for any $g$.
Hence, $\|\dot\Omega_{\theta^*}^{-1}\| < \infty$ and $\dot\Omega_{\theta^*}$ is continuouly invertible.
This verifies condition (F) and completes the verification of all conditions of Theorem 13.4 of \citet{Kosorok:2008}.

\subsubsection*{Step 2}
In this part, we prove that $f_u(z)$ is uniformly bounded away from zero around $\theta_2^*(u)$ and that it is bounded uniformly in $u$ and $x$.
By Assumption~\ref{assumption:bootstrap} (i), $1=\int_K f_{\Psi(Y)}(x)\,dx \leq \overline{f}\cdot \mathrm{Vol}(K)$. Then, we have $\mathrm{Vol}(K)\geq 1/\overline{f}>0$. Combined with the assumption that $K$ is convex, it follows that $\mathrm{int}(K)\neq\emptyset$, so that we can pick $x_0\in\mathrm{int}(K)$ and define
\begin{align*}
    r\coloneqq \sup\{\,\rho>0: B(x_0,\rho)\subset K\,\}>0,\qquad
    R\coloneqq \sup_{x\in K}\|x-x_0\|<\infty,
\end{align*}
where finiteness of $R$ follows from boundedness of $K$.
Then $B(x_0,r)\subset K \subset B(x_0,R)$.
By translating coordinates, we may assume without loss of generality that $x_0=0$, so that $B(0,r)\subset K\subset B(0,R)$.
Here, put $a(u)\coloneqq \min_{x\in K}\langle u,x\rangle$ and $b(u)\coloneqq \max_{x\in K}\langle u,x\rangle$.
Since $K$ is compact and convex, along with the positive density assumption, the support of $Z_u$ is the interval $[a(u),b(u)]$.

We start by showing that the quantile $\theta_2^*(u)$ is uniformly away from the endpoints (that is, it is at an interior point). Recall that $p=\theta_3^*$. Then we can write that
\begin{align*}
    p=\mathbb P(Z_u\geq \theta_2^*(u)\mid S=1,D=1)
    =\int_{K\cap\{\langle u,x\rangle\geq \theta_2^*(u)\}} f_{\Psi(Y)}(x)\,dx \leq 
    \overline{f}\cdot \mathrm{Vol}\left(K\cap\{\langle u,x\rangle\geq \theta_2^*(u)\}\right),
\end{align*}
hence we have $\mathrm{Vol}\left(K\cap\{\langle u,x\rangle\geq \theta_2^*(u)\}\right) \geq p/\overline{f}$.
For $t\in\mathbb R$, let $H_u(t)\coloneqq H^{d-1}\left(K\cap\{x:\langle u,x\rangle=t\}\right)$, where $H^{k}$ is the $k$-dimensional Hausdorff measure.
Fix $u$. Let $h(u)\coloneqq b(u)-\theta_2^*(u) (\geq 0)$. Then we have
\begin{align*}
    \mathrm{Vol}\left(K\cap\{\langle u,x\rangle\geq \theta_2^*(u)\}\right)
    &=\mathrm{Vol}\left(K\cap\{\langle u,x\rangle\ge b(u)-h(u)\}\right)\\
    &=\int_{K\cap\{\langle u,x\rangle\ge b(u)-h(u)\}}1\,d H^d 
    =\int_{b(u)-h(u)}^{b(u)} H_u(t)\,dt,
\end{align*}
where the last equality uses the coarea formula.
Using $K\subset B(0,R)$, any slice satisfies $H_u(t)\leq \kappa_{d-1}R^{d-1}$, where $\kappa_{d-1}$ is the volume of the unit ball in $\mathbb R^{d-1}$. Therefore, we have that $\mathrm{Vol}(K\cap\{\langle u,x\rangle\ge b(u)-h(u)\})\leq h(u)\cdot \kappa_{d-1}R^{d-1}$, implying that
\begin{align*}
    b(u)-\theta_2^*(u)\geq \frac{p}{\overline{f} \kappa_{d-1}R^{d-1}}.
\end{align*}
A similar argument applies to the ``lower bound'' $a(u)$. Hence, there exists a $\delta>0$ such that 
\begin{align*}
    \theta_2^*(u)\in [a(u)+\delta,\ b(u)-\delta],\quad\text{for all }u\in\mathbb S^{d-1}.
\end{align*}
Next, we obtain the uniform lower bound for $H_u(t)$ over an interior region.
Since $B(0,r)\subset K$, we have for every $u$,
\begin{align*}
    H_u(0)\geq H^{d-1}\left(B(0,r)\cap\{x:\langle u,x\rangle=0\}\right)=\kappa_{d-1}r^{d-1}.
\end{align*}
In addition, note that $K_t \coloneqq \left(K\cap\{x:\langle u,x\rangle=t\}\right)$ satisfies $K_{\lambda t_1+(1-\lambda)t_2} \supset \lambda K_{t_1} \oplus (1-\lambda) K_{t_2}$ with $\lambda\in[0,1]$.
Then, combined with the Brunn-Minkowski inequality, for $t\in[0,b(u)]$,
\begin{align*}
    H_u(t)^{1/(d-1)} \geq \frac{b(u)-t}{b(u)} H_u(0)^{1/(d-1)} \geq \frac{b(u)-t}{R} H_u(0)^{1/(d-1)},
\end{align*}
and for $t\in[a(u),0]$,
\begin{align*}
    H_u(t)^{1/(d-1)} \geq \frac{t-a(u)}{-a(u)} H_u(0)^{1/(d-1)} \geq \frac{t-a(u)}{R} H_u(0)^{1/(d-1)}.
\end{align*}
If $t\in[a(u)+\delta/2, b(u)-\delta/2]$, then $\min\{t-a(u),\,b(u)-t\}\geq \delta/2$, so that the above displays imply
\begin{align*}
    H_u(t)\geq
    \left(\frac{\delta}{2R}\right)^{d-1}H_u(0)
    \geq 
    \kappa_{d-1}r^{d-1}\left(\frac{\delta}{2R}\right)^{d-1}.
\end{align*}
Since $u\in\mathbb S^{d-1}$, the coarea formula yields, for $t\in(a(u),b(u))$,
\begin{align*}
    f_u(t)=\int_{K\cap\{x:\langle u,x\rangle=t\}} f_{\Psi(Y)}(x)\,d H^{d-1}(x) \geq \underline{f}\cdot H_u(t),
\end{align*}
which implies that
\begin{align*}
    \inf_{t\in[a(u)+\delta/2, b(u)-\delta/2]}f_u(t) \geq \underline{f}\kappa_{d-1}r^{d-1}\left(\frac{\delta}{2R}\right)^{d-1}\eqqcolon \underline{f}_Z(>0).
\end{align*}
Hence, $\inf_{u\in\mathbb{S}^{d-1}} f_u(\theta_2^*(u))$ and $\inf_{u\in\mathbb{S}^{d-1}} \inf_{t:|t-\theta_2^*(u)|<\eta} f_u(t)$ are bounded away from zero (for some $\eta$ small enough).
A similar argument also shows that $\sup_{u\in\mathbb{S}^{d-1}, x\in \mathbb{R}} f_{u}(x) < \infty$.

\subsubsection*{Step 3}
A similar argument to \textit{Step 1} shows the uniform consistency of $\hat{V}_{\sigma}(u)$.
Then, we have that
\begin{align*}
    \frac{ \sqrt{n}\left(\hat{\sigma}_{\mathcal{I}_1}(u) - {\sigma}_{\mathcal{I}_1}(u)\right)}{\sqrt{\hat{V}_{\sigma}(u)}} 
    &= 
    \frac{ \sqrt{n}\left(\hat{\sigma}_{\mathcal{I}_1}(u) - {\sigma}_{\mathcal{I}_1}(u)\right)}{\sqrt{{V}_\sigma(u)}} + o_\mathbb{P}(1),\\
    \frac{ \sqrt{n}\left(\tilde{\sigma}_{\mathcal{I}_1}(u) - \hat{\sigma}_{\mathcal{I}_1}(u)\right)}{\sqrt{\hat{V}_{\sigma}(u)}}
    &= 
    \frac{ \sqrt{n}\left(\tilde{\sigma}_{\mathcal{I}_1}(u) - \hat{\sigma}_{\mathcal{I}_1}(u)\right)}{\sqrt{{V}_\sigma(u)}} + o_\mathbb{P}(1),\quad
\end{align*}
where $o_\mathbb{P}(1)$ holds uniformly in $u$, which shows that $\hat{\sigma}_{\mathcal{I}_1}(u) + \hat{\text{c.v.}} \sqrt{\hat{V}_{\sigma}(u)/n}\eqqcolon U_n(u)$ is a one-sided uniform confidence band for the support function ${\sigma}_{\mathcal{I}_1}(u)$.
By the property of the support function and convexity of $\hat{\mathcal{R}}_1$, one can easily see that $\{{\sigma}_{\mathcal{I}_1}(u)\leq U_n(u), \forall u\in\mathbb{S}^{d-1}\} = \{\mathcal{S}_1 \subset \hat{\mathcal{R}}_1\}$. This implies that the desired result holds.\hfill$\clubsuit$

\subsection{Proof of Corollary \ref{cor:bootstrap}}
We augment the Z-estimator problem by adding a finite-dimensional components $S(1-D) (\Psi(Y) - \rho)$ to $\psi_{\theta,u}(Y,S,D)$, where $\rho\in\mathbb{R}^d$ is an additional parameter vector with the true value $\rho^*\coloneqq \E{\Psi(Y)\mid S=1,D=0}$. Then, the same argument as in the previous proof shows that $\sqrt{n}(\hat{\theta} - \theta^*, \hat{\rho} - \rho^*)$ and $\sqrt{n}(\tilde{\theta} - \hat{\theta}, \tilde{\rho} - \hat{\rho})$ have the same asymptotic distribution, in the sense that the conditional law of the latter converges in probability to the same limit law as the former. The remainder is also similar to the above.\hfill$\clubsuit$

\bibliographystyle{apalike} 
\bibliography{refs}

\end{document}